\documentclass{article}

\usepackage{arxiv}

\usepackage[utf8]{inputenc} 
\usepackage[T1]{fontenc}    
\usepackage{hyperref}       
\usepackage{url}            
\usepackage{booktabs}       
\usepackage{amsfonts}       
\usepackage{nicefrac}       
\usepackage{microtype}      
\usepackage{lipsum}		
\usepackage{enumitem} 
\usepackage{graphicx}
\usepackage{natbib}
\usepackage{doi}
\usepackage{caption}
\usepackage{multirow}
\usepackage{kotex}

\title{STELLA: Self-Reflective Terminology-Aware Framework for Building an Aerospace Information Retrieval Benchmark}

\author{ Bongmin Kim\\
	TelePIX\\
	\texttt{bmkim@telepix.net} \\
}

\date{}

\renewcommand{\undertitle}{}
\renewcommand{\shorttitle}{STELLA: Framework for Building an Aerospace Information Retrieval Benchmark}

\begin{document}
\maketitle

\begin{abstract}
Tasks in the aerospace industry heavily rely on searching and reusing large volumes of technical documents, yet there is no public information retrieval (IR) benchmark that reflects the terminology- and query-intent characteristics of this domain. To address this gap, this paper proposes the \textbf{STELLA} (\textbf{S}elf-Reflective \textbf{T}\textbf{E}rmino\textbf{L}ogy-Aware Framework for Bui\textbf{L}ding an \textbf{A}erospace Information Retrieval Benchmark) framework. Using this framework, we introduce the \textbf{STELLA benchmark}, an aerospace-specific IR evaluation set constructed from NASA Technical Reports Server (NTRS) documents via a systematic pipeline that comprises document layout detection, passage chunking, terminology dictionary construction, synthetic query generation, and cross-lingual extension. The framework generates two types of queries: the \textit{Terminology Concordant Query} (TCQ), which includes the terminology verbatim to evaluate lexical matching, and the \textit{Terminology Agnostic Query} (TAQ), which utilizes the terminology's description to assess semantic matching. This enables a disentangled evaluation of the lexical and semantic matching capabilities of embedding models. In addition, we combine Chain-of-Density (CoD) and the \textsc{Self-Reflection} method with query generation to improve quality and implement a hybrid cross-lingual extension that reflects real user querying practices. Evaluation of seven embedding models on the STELLA benchmark shows that large decoder-based embedding models exhibit the strongest semantic understanding, while lexical matching methods such as BM25 remain highly competitive in domains where exact lexical matching technical term is crucial. The STELLA benchmark provides a reproducible foundation for reliable performance evaluation and improvement of embedding models in aerospace-domain IR tasks. The STELLA benchmark can be found in \url{https://huggingface.co/datasets/telepix/STELLA}.

\end{abstract}


\section{Introduction}
\label{sec:introduction}
In the aerospace industry, design, manufacturing, and verification tasks heavily rely on searching and reusing large volumes of technical documents. Requirements, design rationales, test conditions, standards compliance, and anomaly analyses are tightly interlinked, and the ability to rapidly locate relevant evidence and draw conclusions is directly tied to productivity and safety. Recently, Retrieval-Augmented Generation (RAG) systems based on Large Language Models (LLMs) have shown the potential to improve the efficiency of such document-centric workflows \citep{lewis2020retrieval, fan2024survey, zhao2024retrieval}. In RAG systems, retrieval augmentation has been reported to reduce hallucinations and improve reliability by grounding the generation process in documents retrieved for a given query \citep{asai2024selfrag, niu2024ragtruth, ayala2024reducing}. However, there is no public information retrieval (IR) benchmark that simultaneously reflects the query intents actually required in the aerospace domain and the lexical and semantic characteristics of domain-specific terminology. Consequently, it is difficult to systematically measure whether an embedding model correctly understands and represents aerospace terminology—namely, whether it captures semantic equivalence beyond mere lexical matching of terms. As a result, it remains challenging to systematically guide embedding model improvement, compare system performance, and assess practical deployability. In practice, BEIR \citep{thakur2021beir} has become a de facto standard for broadly comparing the retrieval performance of embedding models across a variety of public-domain tasks, but it is difficult for such a benchmark to adequately capture and evaluate the lexical and semantic handling of specialized terminology and task-oriented information needs that are central to aerospace-specific retrieval.

Aerospace companies typically rely on internal technical documents and design reports that are not disclosed externally. However, security and intellectual property constraints make it difficult to use such documents directly as public benchmarks. NASA Technical Reports Server (NTRS) documents \citep{nelson1995nasa} broadly share the writing conventions of engineering documents—such as requirement statements, presentation of design rationales, reporting of test and verification results, numerical and unit notation, and references to standards—and are formally and substantively similar to internal technical documents. Although NTRS extensively archives and provides aerospace science and engineering outputs that can serve as a public proxy for internal operational documents, no query–passage relevance set has been established using NTRS as its data source. In other domains, such as medicine and law, domain-specific IR benchmarks have been developed by leveraging large-scale data sources and reflecting user behavior or procedural knowledge \citep{rekabsaz2021tripclick, gao2024enhancing}. In contrast, despite the existence of large-scale source data in the aerospace domain, there is still no well-established evaluation benchmark that can precisely measure which document and which specific evidence a system used to answer a query.

To fill this gap, this paper proposes the \textbf{STELLA} (\textbf{S}elf-Reflective \textbf{T}\textbf{E}rmino\textbf{L}ogy-Aware Framework for Bui\textbf{L}ding an \textbf{A}erospace Information Retrieval Benchmark) framework. The STELLA framework (1) systematically extracts text-centric documents from NTRS aerospace reports, (2) splits the extracted documents into passages that serve as retrieval units in RAG, and (3) precisely extracts aerospace terminology (“terminology”) from the passage set to construct a terminology dictionary. This process applies pattern matching, part-of-speech tagging, and specificity filtering based on sparsity relative to general-purpose corpora to construct a domain-specific terminology dictionary. Next, (4) it performs \textit{Candidate Passage Selection} by using the constructed terminology to select candidate passages that will serve as the passage side of query–passage pairs.  After initially filtering to passages that contain at least five distinct terminology items, the passages are classified into five query intent categories (e.g., \texttt{Definition / Principle}, \texttt{Comparison / Trade-off}) derived in collaboration with domain experts. Then, $k$-medoids clustering is applied within each intent-specific passage pool to extract representative passages for each intent, which are used as candidate passages. Subsequently, (5) synthetic queries are generated based on the representative passages (\textit{Synthetic Query Generation}). The goal of this stage is to construct queries that can measure whether an embedding model effectively understands and represents aerospace terminology. It generates two types of queries: the \textit{Terminology Concordant Query} (TCQ), which includes the terminology verbatim to evaluate lexical matching, and the \textit{Terminology Agnostic Query} (TAQ), which utilizes the description of terminology to assess semantic matching. In this process, the Chain-of-Density (CoD) \citep{adams2023sparse} and \textsc{Self-Reflection} \citep{madaan2023selfrefine, shinn2023reflexion, wang2025critique} methods are applied to the LLM to improve query quality. Finally, (6) a cross-lingual extension is performed to reflect the global collaborative environment of the aerospace industry. TAQ is fully translated into the target language, whereas TCQ is translated in a hybrid translation scheme that preserves terminology in English while translating only the remaining parts, thereby mimicking real user query forms.

In summary, the STELLA framework supports reproducible and practice-oriented improvement of RAG systems through (a) passage construction with domain-consistent aerospace documents, (b) systematic candidate passage selection via terminology extraction and query intent classification, (c) dual-type synthetic query generation that can evaluate both lexical and semantic matching, and (d) cross-lingual extension that reflects real usage patterns. The contributions of this paper are as follows.

\begin{enumerate}
    \item \textbf{STELLA proposal}: We present a domain-specific IR evaluation set construction pipeline, ranging from systematic extraction of text-centric passages from NTRS to synthetic query generation, and we construct and release the resulting STELLA benchmark.
    \item \textbf{Dual-type synthetic query strategy}: We introduce a novel synthetic query generation strategy that can independently measure the lexical matching (TCQ) and semantic matching (TAQ) capabilities of embedding models. CoD and \textsc{Self-Reflection} are applied in this process to generate high-quality queries.
    \item \textbf{Practice-oriented cross-lingual extension}: We define translation rules that preserve English terminology in TCQ translation and perform full translation in TAQ translation to reflect real querying practices, and we provide cross-lingual evaluation sets in six languages.
\end{enumerate}

We expect STELLA to serve as an infrastructure that facilitates reproducible and reliable improvement of aerospace RAG systems, grounded in rigorously constructed, practice-oriented data.

\section{Related Work}
\label{sec:related_work}
\subsection{General IR Benchmarks}
\label{sec:general_ir_benchmarks}
IR benchmarks for objectively evaluating retrieval performance have primarily been developed around open-domain IR datasets. Representative among them, BEIR \citep{thakur2021beir} includes 18 IR tasks spanning diverse domains such as news, Wikipedia, and scientific articles, and has become a de facto standard benchmark for broadly comparing the zero-shot performance of dense and sparse retrievers. Benchmarks such as MTEB~\citep{muennighoff2023mteb} and KILT~\citep{petroni2021kilt} likewise provide extended evaluation frameworks for multi-task learning and knowledge-intensive tasks. However, these generic benchmarks are mostly based on general text documents and thus have limitations in capturing the complex, domain-specific content of specialized fields.

In contrast, the STELLA benchmark constructs passages using materials collected directly from aerospace-domain sources provided by NTRS. As a result, STELLA offers an evaluation set that is conceptually aligned with real operational queries, grounded in domain-consistent aerospace knowledge, terminology, and context. This provides a basis for more precisely revealing domain-specific retrieval difficulty and performance gaps between models that are hard to capture with general-purpose benchmarks.

\subsection{Domain-Specific IR Benchmarks}
\label{sec:domain_specific_ir_benchmarks}
Domain-specific IR benchmarks in fields such as medicine and law have refined query–passage mappings by leveraging large-scale data sources and reflecting document characteristics. \citet{rekabsaz2021tripclick} constructed a large-scale click-based training and evaluation dataset for IR using click logs from a medical search engine. \citet{gao2024enhancing} collected source data from official judgment websites and analyzed the query style of typical users who are unfamiliar with legal knowledge to construct a legal case retrieval dataset.

However, in the aerospace domain, despite the existence of large-scale sources such as NTRS, there is essentially no standardized query–passage relevance set for measuring retriever quality. Existing attempts have mainly focused on downstream tasks such as document-based summarization and question answering, or have covered only subsets of the aerospace domain \citep{emmons2024text, oderinde2025aviation}. In other words, an annotation scheme dedicated to core IR tasks in the aerospace domain has not yet been systematically established. Therefore, this paper is significant in that it proposes an aerospace-specific IR benchmark grounded in systematically constructed guidelines.

\subsection{Multilingual IR Benchmarks}
\label{sec:multilingual_ir_benchmarks}
Given the importance of global collaboration in the aerospace industry, IR systems must robustly handle multilingual queries and documents. Recently, multilingual IR benchmarks such as MIRACL~\cite{zhang2023miracl} have widely evaluated cross-lingual retrieval performance using large-scale open-domain resources centered on Wikipedia, providing meaningful reference points for the language generalization abilities of models \citep{bonifacio2021mmarco, zhang2021mrtydi}. Nevertheless, these benchmarks commonly assume general-domain settings.

By contrast, aerospace-domain documents are predominantly in formalized genres, and their terminology systems and reasoning cues differ in aspects such as abbreviations, standards, component identifiers, and numerical/unit conversions. In addition, information needs are more task-oriented—asking about causes and effects, procedures, and constraints—rather than simple ``fact retrieval''. As a result, models that perform well in cross-lingual open-domain settings may exhibit overestimated or inconsistent performance under domain-specific conditions due to domain shift and failures of normalization \citep{voorhees2021treccovid, liu2023robustness}.

Thus, while multilingual general benchmarks remain a useful starting point, a separate domain-specific cross-lingual IR evaluation set is needed, one that reflects the document structure, terminology systems, and retrieval objectives of the aerospace context. We measure the practical suitability of retrieval models by constructing passages directly from aerospace documents and using task-oriented queries together with a terminology dictionary.

\subsection{Synthetic Query Generation for IR}
\label{sec:synthetic_query_generation_for_ir}
Constructing training and evaluation data for IR models using synthetic queries has recently attracted attention as an efficient data construction approach that leverages the capabilities of LLMs. \citet{bonifacio2022inpars} proposed generating many queries from an LLM and mapping them to positive passages, while \citet{dai2023promptagator} demonstrated that powerful dense retrievers can be trained using queries generated from only a small number of exemplars. \citet{chaudhary2024relative} studied methods for generating synthetic queries using task-specific exemplars and empirically validated their effectiveness. By reconstructing prompts with exemplars related to specific domains instead of general-domain exemplars such as those in MS MARCO \citep{daniel2016msmarco}, they constructed datasets and experimentally showed additional performance gains when training retrievers on these data. This suggests that domain-specific exemplars help align the distribution and style of synthetic queries with the target domain.

Taken together, these studies demonstrate that IR data construction based on LLM knowledge can yield effective datasets without costly manual annotation. We adopt this general passage-to-query synthetic generation paradigm in the aerospace domain. However, relying solely on exemplars has limitations such as (i) omission or bias of key entities within passages, (ii) violations of prohibited formats, and (iii) misclassification of intents and instability of style. To address these issues, we adapt the Chain-of-Density (CoD) concept \citep{adams2023sparse}, which has been effective in summarization, for query generation and perform Self-Refine-style \citep{madaan2023selfrefine, shinn2023reflexion, wang2025critique} \textsc{Self-Reflection} loop at each step. The CoD methodology incrementally identifies and adds key entities of the target text to increase information density, while the \textsc{Self-Reflection} technique is an approach in which an LLM generates feedback on its own outputs during generation and refines them to improve final quality. STELLA applies these two methods to aerospace-domain query generation to systematically recognize key entities such as terminology and to generate precise synthetic queries that reflect complex query intents. Whereas prior work has focused on exemplar-based style alignment, the STELLA framework is distinguished by its prompting strategy that aims to maximize the precision of aerospace-domain-specific queries.

\begin{figure}[t]
    \centering
    \includegraphics[width=\linewidth]{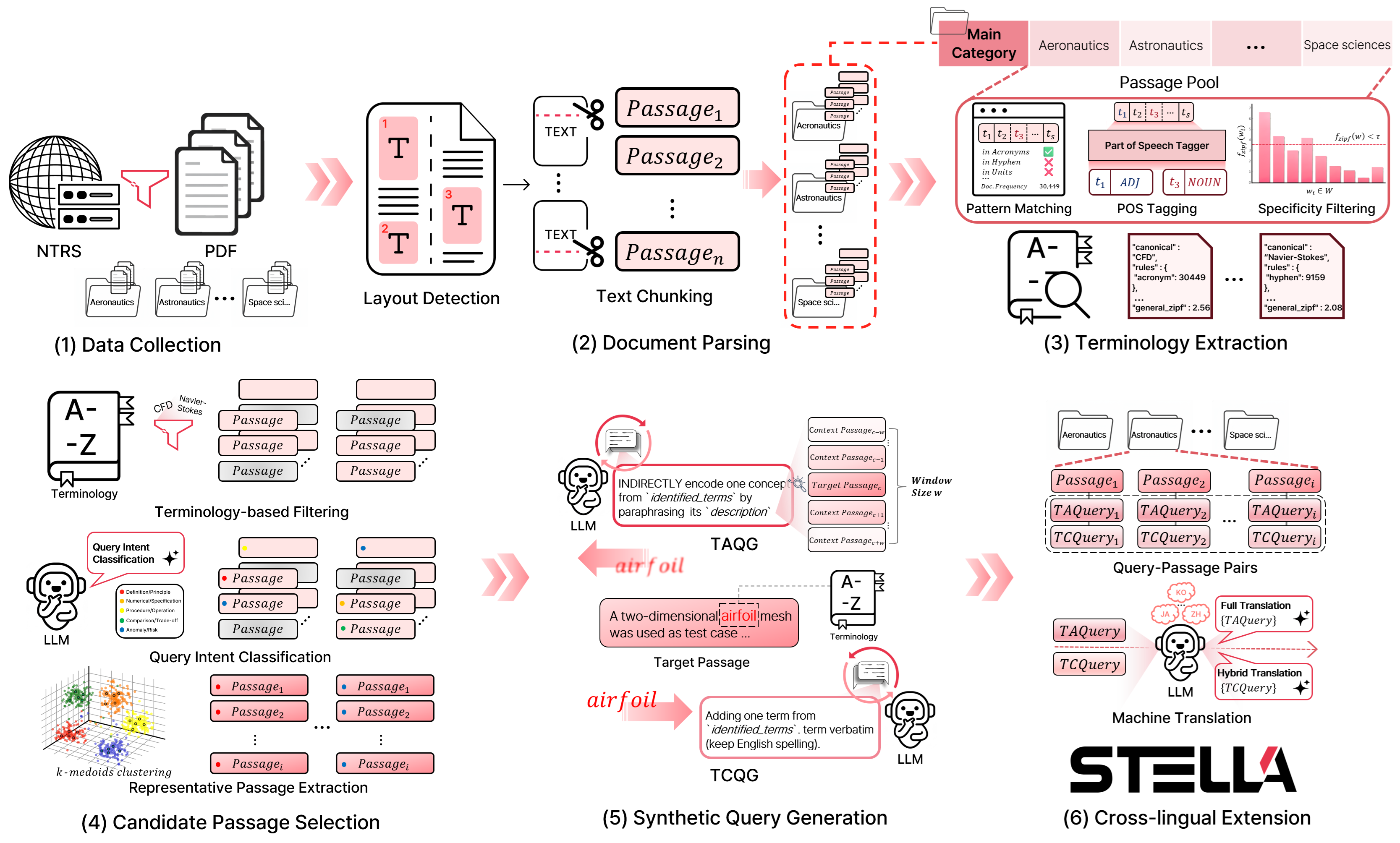}
    \caption{Overall pipeline for constructing the STELLA benchmark. The process comprises six systematic stages: (1) data collection from the NASA Technical Reports Server (NTRS), (2) document layout detection and passage chunking, (3) construction of a domain-specific terminology dictionary, (4) candidate passage selection based on query intents, (5) generation of dual-type synthetic queries (TCQ and TAQ) using Chain-of-Density and \textsc{Self-Reflection}, and (6) cross-lingual extension via multilingual query translation.}
    \label{fig:stella_pipeline}
\end{figure}
\begin{table}[t]
    \centering
    \small
    \begin{tabular}{lrrr}
        \toprule
        Main Category & Total & Usable Amount & Not Usable Amount \\
        \midrule
        Aeronautics                         & 4,372  & 2,860 & 1,512  \\
        Astronautics                        & 2,622  & 1,068 & 1,554  \\
        Chemistry and Materials             & 4,706  & 2,935 & 1,771  \\
        Engineering                         & 5,935  & 3,564 & 2,371  \\
        Geosciences                         & 9,925  & 3,468 & 6,457  \\
        Life Sciences                       & 9,185  & 2,331 & 6,854  \\
        Mathematical and Computer Sciences  & 2,554  & 1,350 & 1,204  \\
        Physics                             & 5,908  & 2,935 & 2,973  \\
        Social and Information Sciences     & 301   & 102  & 199   \\
        Space Sciences                      & 18,405 & 5,864 & 12,541 \\
        \midrule
        \textbf{Total}                      & \textbf{63,913} & \textbf{26,477} & \textbf{37,436} \\
        \bottomrule
    \end{tabular}
    \caption{Statistics of collected NTRS documents by main category. 
    ``Total'' denotes the number of PDF documents whose publication date is 2000 or later (recency criterion), and 
    ``Usable Amount'' denotes the number of documents that additionally satisfy the remaining collection criteria 
    (document format, category selection, and copyright).}
    \label{tab:ntrs_stats_1}
\end{table}
\begin{table}[t]
    \centering
    \small
    \begin{tabular}{lrrrrr}
        \toprule
        Main Category &
        No download URL &
        Duplicate &
        Invalid type &
        Invalid copyright &
        Total excluded \\
        \midrule
        Aeronautics                        &  235 &  211 &  830 &  236 &  1,512 \\
        Astronautics                       &  552 &  139 &  746 &  117 &  1,554 \\
        Chemistry and Materials            &  623 &  162 &  857 &  129 &  1,771 \\
        Engineering                        &  715 &  259 & 1,190 &  207 &  2,371 \\
        Geosciences                        & 3,253 &  274 & 1,517 & 1,413 &  6,457 \\
        Life Sciences                      & 3,752 &  160 & 2,726 &  216 &  6,854 \\
        Mathematical and Computer Sciences &  456 &  127 &  493 &  128 &  1,204 \\
        Physics                            & 1,706 &  189 &  897 &  181 &  2,973 \\
        Social and Information Sciences    &   20 &   30 &  104 &   45 &   199 \\
        Space Sciences                     & 7,150 &  385 & 3,185 & 1,821 & 12,541 \\
        \midrule
        \textbf{Total}                     & \textbf{18,462} & \textbf{1,936} &
                                             \textbf{12,545} & \textbf{4,493} &
                                             \textbf{37,436} \\
        \bottomrule
    \end{tabular}
    \caption{Breakdown of excluded documents by category and filtering criteria described in Section~\ref{sec:source_harvesting}. The columns correspond to the specific exclusion reasons: technical unavailability (No download URL), redundancy handling (Duplicate), format constraints (Invalid type), and licensing restrictions (Invalid copyright). The total exclusions align with the ``Not Usable Amount'' in Table~\ref{tab:ntrs_stats_1}.}
    \label{tab:ntrs_filter_breakdown}
\end{table}

\section{The STELLA Benchmark: Construction Pipeline}
\label{sec:stella_benchmark}
This section details the construction process of the STELLA benchmark, an aerospace-domain-specific IR evaluation set. The STELLA framework consists of a six-stage systematic pipeline: (1) selection and collection of large-scale source documents, (2) document layout detection and passage chunking, (3) terminology dictionary construction, (4) candidate passage selection, (5) synthetic query generation, and (6) dataset extension for cross-lingual evaluation. 
Figure~\ref{fig:stella_pipeline} visualizes the sequential flow of this pipeline and illustrates that the STELLA benchmark is built as a composition of multiple modules rather than a single monolithic task.

\subsection{Data Collection}
\label{sec:source_harvesting}
The corpus underlying the benchmark was collected from NASA Technical Reports Server (NTRS), a reliable public technical resource in the aerospace domain. We aimed to ensure domain consistency by using NTRS documents that share formal and substantive similarities with real operational documents. To ensure data quality and suitability for IR tasks, we selected documents according to the following criteria.

\begin{itemize}
    \item \textbf{Document format}: To account for ease of text extraction and information density, we limited the collection to English-language documents in PDF format and excluded non-text-centric types ``Video'', ``Poster'', ``Presentation'', and ``Abstract''.
    \item \textbf{Recency}: To minimize parsing errors and degraded image quality that may occur in older documents, we included only documents whose publication date is 2000 or later.
    \item \textbf{Category selection}: We set the collection scope to cover all ten main NTRS categories (e.g., ``Aeronautics'', ``Astronautics'') and stored documents according to their category directories. When a document belonged to multiple categories, we used the document identifier (document ID) to remove duplicates and maintain data consistency.
    \item \textbf{Copyright}: To secure materials that can be freely reused and redistributed, we selected only documents that are not under copyright protection and excluded documents without explicit copyright information.
\end{itemize}

For each collected document, we additionally extracted key metadata such as document ID, title, and authors to improve manageability. Statistics of collected documents by category are summarized in Table~\ref{tab:ntrs_stats_1} and Table~\ref{tab:ntrs_filter_breakdown}.

\begin{table}[htbp]
    \centering
    \begin{tabular}{lr}
        \toprule
        \textbf{Category} & \textbf{\# Passages} \\
        \midrule
        Aeronautics & 306,505 \\
        Astronautics & 97,773 \\
        Chemistry and Materials & 225,136 \\
        Engineering & 292,337 \\
        Geosciences & 362,624 \\
        Life Sciences & 197,600 \\
        Mathematical and Computer Sciences & 130,214 \\
        Physics & 242,575 \\
        Social and Information Sciences & 10,516 \\
        Space Sciences & 539,764 \\
        \midrule
        \textbf{Total} & \textbf{2,405,044} \\
        \bottomrule
    \end{tabular}
    \caption{Statistics of passages constructed by category. The text was chunked using the Recursive-Token-Chunker with a size of 100 tokens and an overlap of 20 tokens.}
    \label{tab:passage_stats}
\end{table}

\subsection{Document Parsing}
\label{sec:document_parsing}
We adopted the following approach to extract text from the collected PDF documents and chunk it into passages, which serve as retrieval units. For document layout detection, we used the DocLayout-YOLO \citep{zhao2024doclayout}, which exhibits robust performance across diverse document layouts. DocLayout-YOLO is based on a global-to-local architecture that can detect fine-grained information within documents and is pre-trained on a synthetic dataset of 300K documents, achieving robust layout detection performance across various domains. We explicitly excluded figures and tables that could introduce noise into retrieval process and focused on pure text content, removing bounding boxes whose confidence score was below 0.25 during extraction. Finally, we ordered the bounding boxes according to the natural reading sequence and then extracted the text.

Next, we applied the Recursive-Token-Chunker \citep{langchain2022} to perform chunking, partitioning the extracted text into meaningful units. As reported in \citet{amiri2025chunktwice}, this method better preserves semantic boundaries than fixed-length chunking schemes and thereby improves passage coherence. This is critical for enabling LLMs to clearly understand context and generate high-quality queries. As our chunking strategy, we adopted the configuration that showed optimal performance in multiple experiments reported by \citet{amiri2025chunktwice}, namely a chunk size of 100 tokens with an overlap of 20 tokens (RT100-20). This configuration realistically mimics the chunking environment of actual RAG systems, thereby increasing the likelihood that benchmark results will generalize to real applications. Statistics of passages constructed by category are presented in Table~\ref{tab:passage_stats}.

\subsection{Terminology Extraction}
\label{sec:terminology_extraction}
A key objective of the STELLA benchmark is to assess how well embedding models grasp aerospace terminology, both lexically and semantically. To this end, we construct a high-quality aerospace terminology dictionary that serves as the foundation for the subsequent \textit{Candidate Passage Selection} (Section~\ref{sec:candidate_passage_selection}) and \textit{Synthetic Query Generation} (Section~\ref{sec:synthetic_query_generation}) stages.

The dictionary is built by first broadly extracting candidate technical terms from the entire NTRS corpus and then refining them through multi-stage filtering to retain only highly domain-specific terms. The overall process consists of (1) candidate extraction and (2) multi-stage filtering.

\subsubsection{Candidate Extraction}
\label{sec:candidate_extraction}
First, we extract candidate technical terms from the entire corpus via pattern matching based on regular expressions. This approach is designed to capture common conventions for notating technical terms in aerospace and engineering documents. These include (a) fully capitalized terms (e.g., acronyms such as CFD, MODIS), (b) hyphenated compounds (e.g., ``Navier-Stokes'', ``XMM-Newton''), and (c) terms that include units, Greek letters (e.g., $\alpha$, $\beta$), or mathematical symbols (e.g., ``3-sigma'', $H_2O$).

\subsubsection{Multi-stage Filtering}
\label{sec:multi_stage_filtering}
To ensure relevance and domain specificity, the large set of candidate terms obtained in the first extraction step must satisfy all three of the following criteria to be included in the final dictionary:

\begin{enumerate}[label=\Alph*.]
    \item \textbf{Document frequency}: To remove noise that appears only once in the corpus or arises from parsing errors, we retain only terms that occur in at least ten distinct passages.
    \item \textbf{Part-of-speech tagging}: Noting that most technical terms take nominal forms, we perform part-of-speech tagging using the spaCy library \citep{honnibal2017spacy} and pass to the next stage only those candidate terms identified as nouns or proper nouns.
    \item \textbf{Specificity filtering}: To effectively remove generic words such as ``system'', ``report'', and ``analysis'' which frequently appear in aerospace documents but exhibit low domain specificity, we apply specificity filtering. We utilized the wordfreq \citep{robyn_speer_2022_7199437}, which aggregates word usage statistics from multiple large-scale open-domain corpora, including Wikipedia, Reddit, and Common Crawl, to estimate the general prevalence of term. We compute a general-frequency score $\tau$ for each term and remove terms whose score exceeds a specific threshold ($\tau > 3.5$), i.e., terms that are very common in general-domain usage. Ultimately, only terms satisfying $\tau \le 3.5$ are regarded as aerospace-domain-specific terminology. This threshold is an empirical value chosen to balance domain specificity against coverage. Preliminary analysis confirmed that this threshold provides a trade-off that effectively removes generic words with low domain specificity, such as ``system'' and ``report'' while retaining core technical terms such as ``propellant'' and ``airfoil'' which appear relatively frequently even within the aerospace domain. 
\end{enumerate}

Terms that pass this three-stage filtering process constitute the STELLA terminology dictionary and serve as key resources in candidate passage selection and synthetic query generation.

\subsection{Candidate Passage Selection}
\label{sec:candidate_passage_selection}
The goal of this section is to systematically extract informative and representative passages from the full NTRS passage pool constructed in Section~\ref{sec:document_parsing} that will serve as the basis for \textit{Synthetic Query Generation} (Section~\ref{sec:synthetic_query_generation}). This process comprises three stages: (1) terminology-based filtering, (2) query intent classification, and (3) representative passage extraction.

\subsubsection{Terminology-based Filtering}
\label{sec:terminology_based_filtering}
The first stage performs a coarse selection of passages with high domain specificity—and thus high information value—from the entire NTRS corpus. Using the terminology dictionary built in Section~\ref{sec:terminology_extraction}, we filtered for passages that contained at least five distinct terminology items. This step excludes generic narrative or administrative passages with low technical density and yields a passage pool that focuses on core aerospace concepts.

\subsubsection{Query Intent Classification}
\label{sec:query_intent_classification}
The second stage classifies the once-filtered passages by query intent to reflect the information needs of real users. First, in collaboration with aerospace domain experts, we confirmed that the types of information typically requested from RAG systems in practice can be summarized into five core intents. To incorporate these practical requirements into the benchmark, we performed intent classification on the once-filtered passages using an LLM.

This process was designed so that the LLM acts as a classifier that, given a prompt, determines which type of query each passage is best suited to generate. All LLMs used in this paper are GPT-5~\cite{brown2020language}, and the prompts used for this step are provided in Appendix~\ref{sec:appendix_intent_prompts}. The five query intents are as follows:

\begin{itemize}
    \item \textbf{Definition / Principle (Def)} -- distinguish concepts, explain mechanisms/approximations/variable dependencies.
    \item \textbf{Numerical / Specification (Num)} -- values, ranges, assumptions, units, uncertainties (avoid single-number lookup).
    \item \textbf{Procedure / Operation (Proc)} -- steps, initialization/calibration, schedules, operational rules.
    \item \textbf{Comparison / Trade-off (Comp)} -- quantitative comparisons of performance/mass/power/margins across options/configs.
    \item \textbf{Anomaly / Risk (Anom)} -- causes, reproduction conditions, mitigations/recurrence prevention.
\end{itemize}

Through this step, the filtered passage pool is reorganized into five mutually exclusive intent-specific passage pools.

\subsubsection{Representative Passage Extraction}
\label{sec:representative_passage_extraction}
The final stage extracts passages that best represent each intent from the large intent-specific passage pools. To this end, we first converted all passages in each intent-specific pool into vectors. We used \texttt{embeddinggemma-300m} \citep{vera2025embeddinggemma} as the embedding model. Because this model exhibits the best English embedding performance among models of comparable size on the MTEB, we judged it suitable for effectively capturing semantic relationships between passages.

We then applied $k$-medoids clustering to the embedding pool of each intent. $k$-medoids selects actual data points (medoids) as cluster centers, providing centers that are robust to outliers and interpretable. For each intent-specific pool, we set the number of clusters to $k = 5$. The choice of $k = 5$ was determined as an empirical trade-off in preliminary analysis, capturing diverse sub-topics within each intent while yielding stable clusters.

After clustering, we selected the 20 passages closest to each of the five medoids obtained for each intent-specific pool. This yields $i = 100$ representative passages per intent ($5 \times 20$). In total, 500 passages across the five intents are finalized as candidate passages, which serve as the passage sources in the \textit{Synthetic Query Generation} in Section~\ref{sec:synthetic_query_generation}.

\subsection{Synthetic Query Generation}
\label{sec:synthetic_query_generation}
Based on the candidate passages extracted in Section~\ref{sec:candidate_passage_selection}, we generate synthetic queries using an LLM. The fundamental goal of the STELLA benchmark is to measure how deeply embedding models understand aerospace terminology and its associated context. To achieve this, we generate dual-type queries that allow separate evaluation of two core retrieval capabilities of embedding models.

\begin{enumerate}
    \item \textbf{Terminology Concordant Query (TCQ)}: the terminology appearing in the passage is included verbatim in the query, used to evaluate lexical matching capability.
    \item \textbf{Terminology Agnostic Query (TAQ)}: instead of the terminology itself, the query includes a description or definition of the terminology. This is used to evaluate whether the model can perform conceptual and semantic matching without relying on the surface form of the terminology.
\end{enumerate}

Both types of queries are produced using a prompting strategy that combines the CoD and \textsc{Self-Reflection} methods to control information density and quality during generation.

\subsubsection{Generation Framework for Quality Control: CoD and \textsc{Self-Reflection}}
\label{sec:generation_framework_for_quality_control}
Naive passage-to-query generation tends to produce queries that are ambiguous, overly simplistic, or missing key information from the passage. To overcome these limitations and generate high-quality queries, we applied CoD and \textsc{Self-Reflection}.

\begin{itemize}
    \item \textbf{Chain-of-Density (CoD)}: We adapt the CoD methodology, originally validated in summarization, for query generation \cite{adams2023sparse}. The procedure starts from an initial seed query and progressively increases information density through three steps by adding key entities present in the passage. This approach reduces ambiguity while preventing over-dense queries that include unnecessary information.
    \item \textbf{\textsc{Self-Reflection}}: At each CoD step, we integrate a \textsc{Self-Reflection} mechanism in which the LLM critiques and revises its own queries \citep{madaan2023selfrefine, shinn2023reflexion, wang2025critique}. The LLM checks for violations of the eight hard constraints defined in this work (see Appendix~\ref{sec:appendix_generation_prompts})—for example, ``Is this query answerable solely from the given passage?'', ``Does it contain prohibited formats such as single-number lookup or list requests?'', and ``Does it obey the token length limits while preserving the specified intent?''—and refines its outputs accordingly. This mechanism is key to maintaining the intended purpose and format even in later steps as the queries become more complex.
\end{itemize}

\subsubsection{TCQ and TAQ Generation Procedure}
\label{sec:tcq_and_taq_generation_procedure}
Both types of queries follow the three-step CoD and \textsc{Self-Reflection} framework but differ in how they increase information density.

\paragraph{Terminology Concordant Query Generation (TCQG)}

Starting from the seed query generated in the first step, TCQ is refined in the second and third steps by explicitly adding terminology from the passage (based on the dictionary constructed in Section~\ref{sec:terminology_extraction}). As a result, the TCQs produced in this process exhibit high lexical overlap with the source passages.

\paragraph{Terminology Agnostic Query Generation (TAQG)}

Because TAQ is intended to measure semantic understanding, it strictly prohibits including the terminology itself in the query. Instead, it increases information density by indirectly adding descriptions of the terminology to the seed query from the first step. The descriptions used here are constructed in a preliminary step separate from TAQG.

To generate descriptions for terminology appearing in a specific passage (passage index $c$), we refer to the surrounding context of that passage in the original document. Specifically, we apply a window size of $w = 2$ around the target passage, using five consecutive passages from $c-2$ to $c+2$ as the expanded context passage. The LLM then generates concise descriptions of the terminology based on this expanded context. By using these descriptions as building blocks for TAQG, we encourage models to retrieve passages containing the term ``propellant'' by understanding descriptions such as ``a chemical substance that is burned to propel a rocket'' instead of relying on the word ``propellant'' itself. The specific prompts used for synthetic query generation and terminology description generation are provided in Appendix~\ref{sec:appendix_prompts}. Through the entire process in Section~\ref{sec:synthetic_query_generation}, one TCQ and one TAQ are generated for each of the 500 candidate passages, resulting in an evaluation set of 1,000 unique (query, passage) pairs.

\subsection{Cross-lingual Extension}
\label{sec:cross_lingual_extension}
To reflect the global collaborative nature of the aerospace industry, we extend the benchmark to evaluate cross-lingual retrieval performance. Based on the 1,000 English (query, passage) pairs generated in Section~\ref{sec:synthetic_query_generation}, we translated only the query part into multiple languages. The target languages were selected with typological diversity in mind, covering distinct language families and scripts. The six languages—Korean (ko), Indonesian (id), Thai (th), French (fr), Chinese (zh), and Japanese (ja)—each represent different major language families.

This design enables the STELLA benchmark to robustly evaluate cross-lingual retrieval performance of embedding models across diverse grammatical structures, writing systems, and tokenization schemes. The translation was carried out via prompt learning with an LLM (see Appendix~\ref{sec:appendix_translation_prompts}). However, it is important to go beyond simple machine translation and reflect real querying behavior. Engineers and researchers in the aerospace field tend to retain core technical terminology such as ``RSRM'' or ``propellant'' in English even when searching documents in their native languages.

To accurately incorporate this real-world practice, we applied differentiated translation rules for TAQ and TCQ:

\begin{itemize}
    \item \textbf{TAQ translation}: the entire query is fully translated into the target language. Because TAQ is composed of terminology ``descriptions'' full translation allows us to measure purely semantic retrieval ability in that language environment.
    \item \textbf{TCQ translation}: we perform hybrid translation in which the terminology contained in the query (identified using the dictionary in Section~\ref{sec:terminology_extraction}) is preserved in English while only the descriptive part is translated into the target language. To this end, we carefully designed LLM prompts to enforce terminology preservation (see Table~\ref{tab:prompt_translation}).
\end{itemize}

Finally, including the original English set, all datasets in the six languages of the extended STELLA benchmark are standardized to a schema fully compatible with BEIR, a widely used IR evaluation framework. This enables other researchers to easily use the STELLA benchmark and to reproduce and compare model performance with existing systems. The benchmark is intended to contribute to the advancement of the aerospace IR research ecosystem and to promote reproducible research.

\section{Validation of the STELLA Benchmark}
\label{sec:validation_of_stella_benchmark}
In this section, we verify that each stage of the STELLA benchmark construction pipeline proposed in Section~\ref{sec:stella_benchmark} is statistically and methodologically reliable. These validations provide key evidence supporting the experiments in Section~\ref{sec:experiments}, and we establish the overall soundness of the benchmark by sequentially demonstrating (1) the accuracy of query intent classification, (2) the quality of synthetic queries, and (3) the fidelity of cross-lingual translation.

\begin{table}[t]
    \centering
    \small
    \setlength{\tabcolsep}{7pt}
    \begin{tabular}{llrr}
        \toprule
        \multicolumn{2}{l}{\textbf{Panel A: Overall (5-way)}} &
        \textbf{N} & \textbf{F1} \\
        \midrule
        \multicolumn{2}{l}{Micro-F1} & 300 & 0.933 \\
        \multicolumn{2}{l}{Macro-F1} & 300 & 0.928 \\
        \midrule
        \multicolumn{2}{l}{\textbf{Panel B: Per-intent (expert as reference)}} &
        \textbf{Support} & \textbf{F1} \\
        \midrule
        Def  & Definition / Principle    & 82 & 0.930 \\
        Num  & Numerical / Specification  & 65 & 0.920 \\
        Proc & Procedure / Operation      & 73 & 0.940 \\
        Comp & Comparison / Trade-off     & 51 & 0.910 \\
        Anom & Anomaly / Risk             & 29 & 0.940 \\
        \bottomrule
    \end{tabular}
    \caption{
    F1-based validation of the intent classifier on $N=300$ passages.
    Micro-F1 aggregates over all samples, while Macro-F1 averages F1 across intents to mitigate class-imbalance effects.
    Per-intent F1 is reported with expert-labeled support counts.
    }
    \label{tab:cohens_kappa}
\end{table}

\subsection{Quality Validation of Intent Classification}
\label{sec:reliability_of_query_intent_classification}

We validate the intent classification step in Section~\ref{sec:query_intent_classification}
by measuring how accurately the LLM classifier reproduces expert intent labels.
A domain expert randomly sampled $N=300$ passages from the terminology-filtered pool and assigned
exactly one of the five intents (\texttt{Def}, \texttt{Num}, \texttt{Proc}, \texttt{Comp}, \texttt{Anom}) to each passage.
We then applied the same 300 passages to the LLM classifier and evaluated the predictions using F1-score,
treating the expert labels as the reference.

Because intent frequencies are imbalanced, we report both \textbf{Micro-F1} and \textbf{Macro-F1}.
Micro-F1 reflects overall performance aggregated across all samples, while Macro-F1 averages F1 across intents
and thus highlights performance on minority intents. We further report per-intent F1 with the expert support counts.

Table~\ref{tab:cohens_kappa} shows that the classifier achieves strong overall performance
(Micro-F1 = 0.933, Macro-F1 = 0.928).
Per-intent results indicate consistently high F1 across intents, with relatively lower F1 for
\texttt{Comp} and \texttt{Num}, suggesting these intents are comparatively more confusable under single-label assignment.

\textbf{Limitation.} This validation uses a single expert annotator; therefore, the reported scores measure
fidelity to the expert labels rather than inter-expert reliability. Future work will extend this validation
to multi-expert annotation with adjudication.

\subsection{Quality Validation of Synthetic Queries}
\label{sec:quality_validation_of_synthetic_queries}
We conducted experiments to verify the effectiveness of \textsc{Self-Reflection}, introduced in Section~\ref{sec:synthetic_query_generation} to improve the quality of synthetic queries. To this end, we used eight recent LLMs and compared the quality of generated queries with and without \textsc{Self-Reflection}. For evaluation, we randomly sampled 100 TAQs and 100 TCQs generated by each LLM and scored them with the G-Eval framework \citep{liu2023geval}. Query quality was assessed using the G-Eval framework, with the following five core metrics.

\begin{enumerate}
    \item \textbf{Answerability}: whether the query is answerable solely from the given passage.
    \item \textbf{No External Knowledge}: whether external knowledge is excluded during query generation.
    \item \textbf{Intent Adherence}: whether the predefined query intent is respected.
    \item \textbf{Format Compliance}: whether prohibited formats (e.g., list requests, single-word answers) are avoided.
    \item \textbf{Style \& Length}: whether the specified style (neutral, technical) and length constraints are satisfied.
\end{enumerate}

Each metric was scored on a 1–5 scale, and the final score was computed as the average of the five metrics. The models under evaluation are grouped into two categories according to their parameter scale.

\begin{itemize}
    \item \textbf{Large-scale models (over 200B or mixture-of-experts)}: GPT-5~\cite{brown2020language}, DeepSeek-V3.2-Exp~\cite{deepseekai2025v32exp}, Qwen3-235B-A22B-Instruct~\cite{yang2025qwen3}, and Llama-4-Maverick~\cite{meta2025llama4}. These are state-of-the-art flagship models with extensive knowledge and advanced reasoning capabilities.
    \item \textbf{Small-scale models (7B–8B dense)}: Qwen3-8B~\cite{yang2025qwen3}, Llama-3.1-8B-Instruct~\cite{meta2024llama3herd}, Mistral-7B-Instruct-v0.3~\cite{jiang2023mistral7b}, and DeepSeek-R1-0528-Qwen3-8B~\cite{deepseekai2025deepseekr1}. These models are designed with efficiency in mind for resource-constrained environments.
\end{itemize}

\begin{figure*}[t!]
    \centering
    \includegraphics[width=0.95\textwidth]{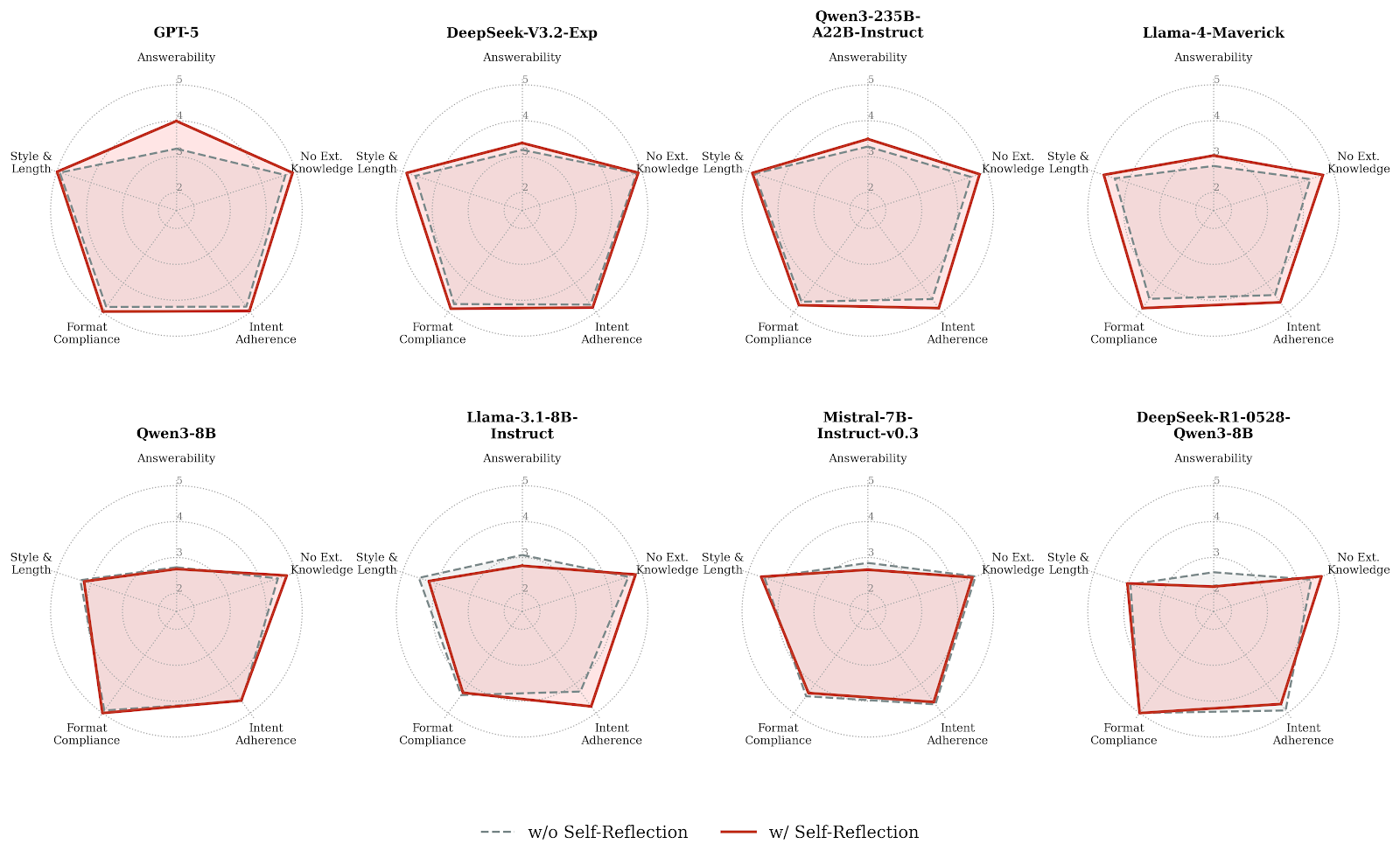}
    \caption{
        Quality Validation of Synthetic Queries via G-Eval.
        Performance comparison of synthetic query generation with (red solid line) and without (grey dashed line) \textsc{Self-Reflection} across five core metrics: \texttt{Answerability}, \texttt{No External Knowledge}, \texttt{Intent Adherence}, \texttt{Format Compliance}, and \texttt{Style \& Length}. 
        The top row presents large-scale models, showing significant improvement in \texttt{Answerability} and constraint adherence. 
        The bottom row presents small-scale models, where the benefits of \textsc{Self-Reflection} are limited or negative due to reasoning limitations.
    }
    \label{fig:quality_validation_radar}
\end{figure*}

\textbf{Effectiveness on large-scale models.} As shown in Figure~\ref{fig:quality_validation_radar}, the radar plots for the \textsc{Self-Reflection} setting expand outward along all axes compared to the setting without \textsc{Self-Reflection}. In particular, GPT-5 exhibits the most pronounced change on the \texttt{Answerability} axis, increasing substantially from 3.22 to 3.99 (+0.77). This suggests that the \textsc{Self-Reflection} process regarding whether the generated query is answerable solely from the passage is highly effective. Furthermore, the fact that the \texttt{Format Compliance} and \texttt{No External Knowledge} axes approach the maximum score of 5.0 for all models indicates that \textsc{Self-Reflection} plays a crucial role in enforcing the hard constraints.

\textbf{Limitations on small-scale models.} By contrast, in the small-scale model group, the effect of \textsc{Self-Reflection} was minor or even negative. On average across the four small models, \texttt{Answerability} scores decreased when \textsc{Self-Reflection} was applied, and the overall mean score also showed a slight decline. We attribute this to the limited reasoning capacity of small models. \textsc{Self-Reflection} requires models to critically evaluate and revise their own outputs, a complex task that small models may struggle with. In some cases, erroneous \textsc{Self-Reflection} appears to degrade the quality of initially reasonable generations. An exception is Qwen3-8B, which showed a slight performance improvement, although the gain remains modest compared to large models.

Based on these findings, the synthetic queries in this benchmark were ultimately constructed using GPT-5 with \textsc{Self-Reflection}, which provided the highest quality.

\subsection{Quality Validation of Cross-lingual Translation}
\label{sec:quality_validation_of_cross_lingual_translation}
We conducted two validation procedures to ensure the quality of the LLM-based translation described in Section~\ref{sec:cross_lingual_extension}. First, we quantitatively evaluated semantic fidelity via back-translation. After translating all queries into six languages and then back into English, we measured the cosine similarity between embeddings of the original and back-translated queries, obtaining high average scores of 0.93 or higher for all languages, which confirms that meaning is well preserved during translation.

Second, we automatically inspected all translated TCQs to verify compliance with the hybrid translation rules. We confirmed that all queries accurately preserve English terminology, demonstrating that the benchmark successfully emulates real user query forms.

\section{Experiments}
\label{sec:experiments}
\subsection{Experimental Setup}
\label{sec:experimental_setup}
To validate the usefulness of the STELLA benchmark and analyze retrieval performance in the aerospace domain from multiple perspectives, we evaluated one lexical baseline and seven recent neural embedding models. The models were selected along the following three key axes:

\begin{enumerate}
    \item \textbf{Architecture:} We compare traditional encoder-only bi-encoder models with recent decoder-only LLM-derived models.
    \item \textbf{Model Scale:} We analyze the trade-off between knowledge capacity and efficiency for lightweight models with 0.3B–0.6B parameters and large models with 7B–8B parameters.
    \item \textbf{Multilingual Capability:} We consider whether each model supports multilingual environments to align with the cross-lingual extension in Section~\ref{sec:cross_lingual_extension}.
\end{enumerate}

Table~\ref{tab:model_summary} provides a comprehensive summary of the evaluated models. The evaluated models are categorized by architectural paradigm into a lexical baseline and two groups of neural embedding models.
\begin{table}[t]
    \centering
    \small
    \begin{tabular}{l c c c}
        \toprule
        \textbf{Model} & \textbf{Architecture} & \textbf{Params} & \textbf{Ref.} \\
        \midrule
        \multicolumn{4}{l}{\textit{Lexical Baseline}} \\
        BM25 & Probabilistic & N/A & \cite{robertson1995okapi} \\
        \midrule
        \multicolumn{4}{l}{\textit{Encoder-only (Group 1)}} \\
        Arctic-Embed-2.0-L & Bi-encoder & 0.6B & \cite{yu2024arctic} \\
        BGE-M3 & Bi-encoder & 0.6B & \cite{chen2024m3embedding} \\
        mE5-instruct & Bi-encoder & 0.6B & \cite{wang2024multilingual} \\
        mGTE & Bi-encoder & 0.3B & \cite{zhang2024mgte} \\
        \midrule
        \multicolumn{4}{l}{\textit{Decoder-only (Group 2)}} \\
        Llama-Embed-Nemotron & LLM Decoder & 8B & \cite{babakhin2025llamaembed} \\
        Qwen3-Embedding & LLM Decoder & 8B & \cite{zhang2025qwen3embedding} \\
        SFR-Embedding-Mistral & LLM Decoder & 7B & \cite{meng2024sfr} \\
        \bottomrule
    \end{tabular}
    \caption{Summary of Evaluated Models}
    \label{tab:model_summary}
\end{table}

\textbf{Lexical baseline.}

\begin{itemize}
    \item \textbf{BM25}: a classical probabilistic IR ranking function that operates on lexical matching rather than semantic similarity. In our experiments, BM25 serves as a strong baseline for evaluating both TCQ and TAQ performance as defined in Section~\ref{sec:synthetic_query_generation} and for comparing against the semantic retrieval performance of neural embedding models~\cite{robertson1995okapi}.
\end{itemize}

\textbf{Group 1: encoder-only architectures.}
These models follow the traditional bi-encoder paradigm and are characterized by relatively small size and high inference efficiency.

\begin{itemize}
    \item \textbf{Arctic-Embed-2.0-L (0.6B)}: fine-tuned from \texttt{bge-m3-retromae} and designed to balance high retrieval performance with inference efficiency~\cite{yu2024arctic}.
    \item \textbf{BGE-M3 (0.6B)}: developed by fine-tuning an XLM-RoBERTa~\cite{conneau-etal-2020-unsupervised} and characterized by multi-functionality~\cite{chen2024m3embedding}.
    \item \textbf{mE5-instruct (0.6B)}: a multilingual instruction-tuned model derived from the E5 series~\cite{wang2024multilingual}.
    \item \textbf{mGTE (0.3B)}: part of the GTE model series, designed to provide efficient yet strong retrieval performance in multilingual settings~\cite{zhang2024mgte}.
\end{itemize}

\textbf{Group 2: decoder-only architectures.}  
These are recent models that apply LLMs with more than 7B parameters to embedding tasks and are characterized by strong semantic understanding grounded in extensive pre-training.

\begin{itemize}
    \item \textbf{Llama-Embed-Nemotron (8B)}: based on \texttt{Llama-3.1-8B} and replacing the inherent unidirectional attention of the decoder with bi-directional self-attention, enabling richer semantic understanding over the full token context~\cite{babakhin2025llamaembed}.
    \item \textbf{Qwen3-Embedding (8B)}: an embedding model from the Qwen3 series that, unlike Llama-Embed-Nemotron, retains the core decoder architecture while optimizing embedding performance~\cite{zhang2025qwen3embedding}.
    \item \textbf{SFR-Embedding-Mistral (7B)}: based on the \texttt{Mistral-7B-v0.1}~\cite{jiang2023mistral7b} architecture and evaluated alongside Qwen3-Embedding to assess the embedding performance of LLM decoder architectures~\cite{meng2024sfr}.
\end{itemize}
We measure model retrieval performance using nDCG@k (normalized discounted cumulative gain at k), a standard evaluation metric in IR. nDCG@k jointly accounts for the relevance and ranking of the top-$k$ retrieved results. In our experiments, we set $k = 10$ as the default and focus on the quality of the top ten results (nDCG@10), which is consistent with typical user scenarios.

\begin{table*}[t]
    \centering
    \tiny
    \begin{tabular}{l cc | cccccc | cccccc}
        \toprule
        \multirow{2}{*}{\textbf{Model}} & \multirow{2}{*}{\textbf{Overall}} & \multirow{2}{*}{\textbf{Gap}} & \multicolumn{6}{c}{\textbf{Terminology Concordant Query (TCQ)}} & \multicolumn{6}{c}{\textbf{Terminology Agnostic Query (TAQ)}} \\
        \cmidrule(lr){4-9} \cmidrule(lr){10-15}
         & & & \textbf{Anom} & \textbf{Comp} & \textbf{Def} & \textbf{Num} & \textbf{Proc} & \textbf{Avg} & \textbf{Anom} & \textbf{Comp} & \textbf{Def} & \textbf{Num} & \textbf{Proc} & \textbf{Avg} \\
        \midrule
        \multicolumn{15}{l}{\textit{Lexical Baseline}} \\
        BM25 & 0.659 & 0.228 & \underline{0.871} & \underline{0.858} & 0.572 & 0.677 & \underline{0.886} & 0.773 & 0.616 & 0.609 & 0.458 & 0.434 & 0.609 & 0.545 \\
        \midrule
        \multicolumn{15}{l}{\textit{Encoder-only (Group 1)}} \\
        Arctic-Embed-2.0-L & 0.672 & 0.227 & 0.870 & 0.839 & 0.598 & \underline{0.765} & 0.854 & \underline{0.785} & 0.639 & 0.611 & 0.438 & 0.496 & 0.608 & 0.558 \\
        BGE-M3 & 0.628 & 0.272 & 0.808 & 0.841 & 0.563 & 0.761 & 0.846 & 0.764 & 0.596 & 0.476 & 0.353 & 0.515 & 0.520 & 0.492 \\
        mE5-instruct & 0.476 & 0.230 & 0.628 & 0.646 & 0.442 & 0.581 & 0.656 & 0.591 & 0.409 & 0.382 & 0.287 & 0.314 & 0.412 & 0.361 \\
        mGTE & 0.572 & 0.220 & 0.739 & 0.751 & 0.490 & 0.664 & 0.768 & 0.682 & 0.551 & 0.517 & 0.310 & 0.447 & 0.487 & 0.462 \\
        \midrule
        \multicolumn{15}{l}{\textit{Decoder-only (Group 2)}} \\
        Llama-Embed-Nemotron & \textbf{0.788} & \textbf{0.106} & \textbf{0.872} & \textbf{0.887} & \textbf{0.684} & \textbf{0.872} & \textbf{0.888} & \textbf{0.841} & \textbf{0.792} & \textbf{0.810} & \textbf{0.532} & \textbf{0.753} & \textbf{0.787} & \textbf{0.735} \\
        Qwen3-Embedding & \underline{0.694} & \underline{0.171} & 0.818 & 0.812 & \underline{0.644} & 0.753 & 0.868 & 0.779 & \underline{0.668} & 0.637 & \underline{0.528} & 0.520 & \underline{0.688} & \underline{0.608} \\
        SFR-Embedding-Mistral & 0.660 & 0.183 & 0.799 & 0.844 & 0.555 & 0.760 & 0.798 & 0.751 & 0.613 & \underline{0.657} & 0.373 & \underline{0.547} & 0.651 & 0.568 \\
        \bottomrule
    \end{tabular}
        \caption{
    Comprehensive retrieval performance on the English subset of the STELLA benchmark.
    This table presents the \textbf{Overall} nDCG@10 scores and the performance \textbf{Gap} between Terminology Concordant Query (TCQ) and Terminology Agnostic Query (TAQ), where a lower gap indicates reduced lexical dependency. 
    Detailed performance breakdowns are provided for five specific query intents under both TCQ and TAQ settings. 
    The best results are \textbf{bolded}, and the second-best are \underline{underlined}.}
    
    \label{tab:main_results}
    \vspace{0.5ex}

\end{table*}

\subsection{Overall Performance Comparison}
\label{sec:overall_preformance_comparison}
The overall performance on the English subset of the STELLA benchmark is summarized in Table~\ref{tab:main_results}. Before analyzing the cross-lingual capabilities, we first establish a baseline for domain-specific retrieval performance in a monolingual setting. This shows that recent neural models do not always dominate in specialized domains such as aerospace. Llama-Embed-Nemotron achieves a substantial lead over other models, demonstrating the advantage of its architecture and scale. The competition among the remaining models is more nuanced. The Arctic-Embed-2.0-L (0.6B) exhibits impressive performance, slightly surpassing both the SFR-Embedding-Mistral (7B) and the BM25.

This suggests that domain suitability and the effectiveness of training objectives can be more decisive for performance than model size alone. In contrast, models such as mE5-instruct and mGTE, which aim for broad general performance, fall short of the BM25 baseline, revealing their limitations in highly specialized domains.

\subsection{Impact of Terminology on TCQ vs. TAQ}
\label{sec:impact_of_terminology}
Analyzing performance differences with and without terminology clearly reveals the characteristics of each model. The lexical baseline BM25 performs strongly on TCQ, where technical terms appear explicitly, but its performance drops sharply on TAQ, yielding a large lexical dependency gap of about 0.228 (TCQ$-$TAQ).

Among neural embedding models, Llama-Embed-Nemotron minimizes this gap to around 0.1053, demonstrating strong conceptual understanding that captures contextual meaning even when surface technical terms are absent. By contrast, BGE-M3 exhibits an even larger gap (0.272) than BM25, indicating that it behaves more like a keyword matcher than a model capturing deep semantic connections in specialized domains.

\subsection{Analysis by Query Intent}
\label{sec:analysis_by_query_intent}
By query intent, models generally perform well on \texttt{Comparison / Trade-off} and \texttt{Anomaly / Risk}. Notably, BM25 achieves a very high score of 0.886 for \texttt{Procedure / Operation} under the TCQ. We attribute this to the nature of procedural documents, in which certain technical terms are repeatedly used, making lexical retrieval particularly effective.

The most challenging intent is \texttt{Definition / Principle}, where all models struggle, especially under TAQ. Even Llama-Embed-Nemotron attains only 0.532 in this setting, indicating that retrieving abstract principles without terminology matching remains difficult even for state-of-the-art models.

\subsection{Architectural Insights}
\label{sec:architectural_insigh}
These experimental results yield several key insights for designing domain-specific embedding models.

\textbf{Need for large decoder-based models.} The experiments show that architectures with small lexical dependency gaps—and thus better conceptual understanding—are large decoder-based models. In particular, the only model that clearly surpasses the strong BM25 baseline is Llama-Embed-Nemotron, which incorporates bi-directional attention. The poorer performance of similarly sized unidirectional models (Qwen3-Embedding, SFR-Embedding-Mistral) compared with Llama-Embed-Nemotron suggests that both model scale and bi-directional contextual understanding are required for domain-specific retrieval.

\textbf{Efficiency of encoders.} The fact that the Arctic-Embed-2.0-L (0.6B) performs on par with the SFR-Embedding-Mistral (7B)—which is over ten times larger—highlights the efficiency of encoder architectures. This result indicates that neither model size nor performance on general benchmarks necessarily guarantees strong performance in real-world or domain-specific environments, underscoring the importance of domain-appropriate experiments and evaluation. It also suggests that models like Arctic-Embed-2.0-L can be attractive alternatives in resource-constrained settings.

\textbf{Reaffirming the strength of lexical baselines.} The fact that many recent neural models fall short of BM25 shows that lexical matching remains highly competitive in domains where discrimination over specialized terminology is crucial. Therefore, hybrid approaches should be considered essential when building practical systems.

\begin{table*}[t]
    \centering
    \tiny
    \setlength{\tabcolsep}{5.5pt} 
    \begin{tabular}{l c c | cccccc c | cccccc c}
        \toprule
        \multirow{2}{*}{\textbf{Model}} & \textbf{Ref.} & \textbf{All} & \multicolumn{7}{c|}{\textbf{Terminology Concordant Query (TCQ)}} & \multicolumn{7}{c}{\textbf{Terminology Agnostic Query (TAQ)}} \\
        \cmidrule(lr){2-2} \cmidrule(lr){3-3} \cmidrule(lr){4-10} \cmidrule(lr){11-17}
         & \textbf{en} & \textbf{Avg} & \textbf{ko} & \textbf{id} & \textbf{th} & \textbf{fr} & \textbf{zh} & \textbf{ja} & \textbf{Avg} & \textbf{ko} & \textbf{id} & \textbf{th} & \textbf{fr} & \textbf{zh} & \textbf{ja} & \textbf{Avg} \\
        \midrule
        \multicolumn{17}{l}{\textit{Encoder-only (Group 1)}} \\
        Arctic-Embed-2.0-L & 0.672 & 0.524 & 0.661 & \underline{0.717} & 0.643 & \underline{0.722} & 0.677 & 0.698 & 0.686 & 0.345 & 0.422 & 0.290 & 0.402 & 0.317 & 0.393 & 0.362 \\
        BGE-M3 & 0.628 & 0.485 & 0.635 & 0.680 & 0.650 & 0.673 & 0.627 & 0.667 & 0.655 & 0.290 & 0.356 & 0.286 & 0.332 & 0.292 & 0.335 & 0.315 \\
        mE5-instruct & 0.476 & 0.199 & 0.292 & 0.393 & 0.367 & 0.423 & 0.156 & 0.233 & 0.311 & 0.075 & 0.131 & 0.086 & 0.161 & 0.022 & 0.046 & 0.087 \\
        mGTE & 0.572 & 0.383 & 0.475 & 0.546 & 0.461 & 0.554 & 0.584 & 0.523 & 0.524 & 0.196 & 0.288 & 0.165 & 0.310 & 0.271 & 0.219 & 0.242 \\
        \midrule
        \multicolumn{17}{l}{\textit{Decoder-only (Group 2)}} \\
        Llama-Embed-Nemotron & \textbf{0.788} & \textbf{0.707} & \textbf{0.798} & \textbf{0.818} & \textbf{0.774} & \textbf{0.806} & \textbf{0.801} & \textbf{0.811} & \textbf{0.801} & \textbf{0.618} & \textbf{0.632} & \textbf{0.529} & \textbf{0.634} & \textbf{0.625} & \textbf{0.633} & \textbf{0.612} \\
        Qwen3-Embedding & \underline{0.694} & \underline{0.603} & \underline{0.718} & 0.696 & \underline{0.675} & 0.702 & \underline{0.725} & \underline{0.720} & \underline{0.706} & \underline{0.484} & \underline{0.508} & \underline{0.432} & \underline{0.528} & \underline{0.524} & \underline{0.519} & \underline{0.499} \\
        SFR-Embedding-Mistral & 0.660 & 0.424 & 0.548 & 0.616 & 0.450 & 0.656 & 0.566 & 0.593 & 0.572 & 0.238 & 0.322 & 0.139 & 0.421 & 0.275 & 0.261 & 0.276 \\
        \bottomrule
    \end{tabular}
    \vspace{0.5ex}
    \caption{
        Detailed cross-lingual retrieval performance.
        \textbf{Ref.} denotes monolingual performance from Table~\ref{tab:main_results}.
        \textbf{All} represents the overall average across all cross-lingual settings.
        For \textbf{TCQ} and \textbf{TAQ}, the detailed scores for six languages are followed by their respective averages (\textbf{Avg}) at the far right.
        Best results are \textbf{bolded}, second-best are \underline{underlined}.
    }
    \label{tab:cross_lingual}
\end{table*}

\subsection{Cross-lingual Performance}
\label{sec:cross_lingual_performance}
To reflect the global collaborative environment of the aerospace industry, we evaluated cross-lingual retrieval performance using queries translated into the six languages defined in Section~\ref{sec:cross_lingual_extension} (Table~\ref{tab:cross_lingual}). This experiment measures the ability to match multilingual queries to English passages.

\textbf{Overall performance.} The relative ordering of models observed in the monolingual setting is largely preserved in the cross-lingual setting, but all models experience performance degradation. Llama-Embed-Nemotron remains the top-performing model by a wide margin in the cross-lingual setting. It shows the smallest drop in performance, about 8.1\% relative to the monolingual baseline, demonstrating strong cross-lingual generalization. Qwen3-Embedding also attains an average of 0.603 across the six languages, corresponding to a 9.1\% drop from monolingual setting.

In contrast, encoder models that emphasize multilingual support exhibit substantial degradation. BGE-M3 drops by about 14.3\% relative to monolingual setting, and Arctic-Embed-2.0-L by about 14.8\%. mE5-instruct records an average of 0.199 across the six languages, a sharp 27.7\% decrease from the monolingual setting, indicating that it is largely ineffective for cross-lingual retrieval. In contrast, the mGTE model, with roughly half as many parameters, shows only a 18.9\% decrease in the cross-lingual setting, exhibiting robust retrieval performance.

\textbf{Language-wise performance bias and weaknesses.} The linguistic diversity in the benchmark reveals clear weaknesses for each model. Thai (th) yields the largest performance drop for most models. Its unique script and grammatical structure make Thai the most challenging language, with Llama-Embed-Nemotron also exhibiting a notably larger drop compared with other languages. mE5-instruct shows severe weaknesses especially for Asian languages, recording the lowest performance in Japanese and Chinese. Interestingly, SFR-Embedding-Mistral shows strength in French (fr), achieving 0.539, but suffers sharp drops in Thai and Korean, revealing linguistic biases.

\textbf{Impact of terminology.} The cross-lingual experiments clearly validate the effectiveness of the translation rules designed in Section~\ref{sec:cross_lingual_extension}. For all models, TCQ performance—where  terminology is preserved in English—is substantially higher than TAQ performance, which is fully translated. This is because keeping technical terms in their original English form perfectly circumvents the difficult problem of cross-lingual alignment of domain-specific knowledge that multilingual models have yet to solve, and our results provide clear empirical evidence that the ``hybrid queries'' issued by real-world users, as assumed in Section~\ref{sec:cross_lingual_extension}, are a highly reasonable strategy.

The gap is especially pronounced for encoder models such as BGE-M3. It achieves an average TCQ score of 0.655 across the six languages, but its TAQ score drops to 0.315, less than half. As in the monolingual analysis in Section~\ref{sec:impact_of_terminology}, this shows that the model still strongly depends on surface-form term matching. Even the top-performing model Llama-Embed-Nemotron exhibits a large gap of about 0.189 between its average TCQ and TAQ scores. This gap is larger than that observed in the monolingual setting, indicating that simultaneously overcoming both conceptual understanding and language barriers remains a very challenging task for current models.

\section{Conclusion}
\label{sec:conclusion}
This paper proposes the STELLA framework to address the lack of practical evaluation of IR performance in the aerospace domain and introduces the resulting benchmark. The STELLA framework is centered on dual-type queries that separately measure lexical matching (TCQ) and semantic matching (TAQ), quality control via \textsc{Self-Reflection}, and a hybrid cross-lingual extension that reflects real usage patterns.
After validating the benchmark’s reliability and evaluating seven embedding models, we confirmed that TAQ constitutes the most challenging setting for current neural embedding models. Moreover, the strong performance of the BM25 baseline reaffirms the need for hybrid retrieval, and the practical utility of TCQ with preserved terminology is demonstrated even in cross-lingual settings.
In conclusion, the STELLA benchmark provides a core evaluation foundation for precisely identifying weaknesses of embedding models used in aerospace RAG systems, and the proposed framework fosters future research on domain-specific model development and hybrid retrieval strategies.

\section{Limitations and Future Work}
\label{sec:limitations_and_future_work}
STELLA has several inherent limitations. First, because the benchmark corpus relies on NTRS as a single public repository, it may not fully represent proprietary documents in industrial environments or materials from other institutions. Second, we intentionally excluded non-textual information such as figures and tables during construction, so the benchmark cannot evaluate retrieval capabilities over structured or multimodal data beyond text. Third, because all query–passage pairs are synthesized via passage-to-query generation with an LLM, it is difficult to fully emulate the diversity of real user queries that may be ambiguous, require combining multiple documents (multi-hop), or be unanswerable. Fourth, heuristic assumptions such as the “at least five terms” filter and the five intent categories are embedded throughout the construction pipeline and may oversimplify the complexity of real-world information needs. Finally, the cross-lingual extension relies on LLM-based translation, and the hybrid translation rule of preserving terminology in TCQ, while reflecting practical conventions, also limits coverage of diverse user behaviors that translate or transliterate terms into their native languages.

Although this work is a first step toward filling the evaluation gap in aerospace-domain IR, several extensions are needed. First, we plan to expand the complexity and scope of the benchmark. Beyond text-only corpora, we aim to extend it to a multimodal IR evaluation set that includes figures, tables, and graphs from real technical documents. At the same time, we will enhance the realism of the benchmark by adding multi-hop queries that require referencing multiple documents and unanswerable query types in addition to single-passage synthetic queries. Second, we seek to increase corpus and language diversity. We plan to incorporate public technical reports and patent documents from institutions other than NTRS to reduce data bias. In addition, beyond the TCQ terminology-preservation rule in Section~\ref{sec:cross_lingual_extension}, we will diversify cross-lingual evaluation scenarios by reflecting real user behaviors that fully translate or transliterate terminology into their native languages.

\bibliographystyle{unsrtnat}
\bibliography{references}  

@inproceedings{lewis2020retrieval,
    title = {Retrieval-Augmented Generation for Knowledge-Intensive {NLP} Tasks},
    author = {Lewis, Patrick and Perez, Ethan and Piktus, Aleksandra and Petroni, Fabio and Karpukhin, Vladimir and Goyal, Naman and K{\"u}ttler, Heinrich and Lewis, Mike and Yih, Wen-tau and Rockt{\"a}schel, Tim and Riedel, Sebastian and Kiela, Douwe},
    booktitle = {Advances in Neural Information Processing Systems},
    volume = {33},
    pages = {9459--9474},
    year = {2020},
}

@inproceedings{brown2020language,
    title = {Language Models are Few-Shot Learners},
    author = {Brown, Tom B. and Mann, Benjamin and Ryder, Nick and Subbiah, Melanie and Kaplan, Jared and Dhariwal, Prafulla and Neelakantan, Arvind and Shyam, Pranav and Sastry, Girish and Askell, Amanda and Agarwal, Sandhini and Herbert-Voss, Ariel and Krueger, Gretchen and Henighan, Tom and Child, Rewon and Ramesh, Aditya and Ziegler, Daniel M. and Wu, Jeffrey and Winter, Clemens and Hesse, Christopher and Chen, Mark and Sigler, Eric and Litwin, Mateusz and Gray, Scott and Chess, Benjamin and Clark, Jack and Berner, Christopher and McCandlish, Sam and Radford, Alec and Sutskever, Ilya and Amodei, Dario},
    booktitle = {Advances in Neural Information Processing Systems},
    volume = {33},
    pages = {1877--1901},
    year = {2020},
}

@inproceedings{fan2024survey,
    title = {A Survey on {RAG} Meeting {LLMs}: Towards Retrieval-Augmented Large Language Models},
    author = {Fan, Wenqi and Ding, Yujuan and Ning, Liangbo and Wang, Shijie and Li, Hengyun and Yin, Dawei and Chua, Tat-Seng and Li, Qing},
    booktitle = {Proceedings of the 30th ACM SIGKDD Conference on Knowledge Discovery and Data Mining},
    year = {2024},
    doi = {10.1145/3637528.3671470},
}

@article{zhao2024retrieval,
    title = {Retrieval-Augmented Generation for {AI}-Generated Content: A Survey},
    author = {Zhao, Penghao and Zhang, Hailin and Yu, Qinhan and Wang, Zhengren and Geng, Yunteng and Fu, Fangcheng and Yang, Ling and Zhang, Wentao and Jiang, Jie and Cui, Bin},
    journal = {arXiv preprint arXiv:2402.19473},
    year = {2024},
}

@inproceedings{asai2024selfrag,
    title = {{Self-RAG}: Learning to Retrieve, Generate, and Critique through Self-Reflection},
    author = {Asai, Akari and Wu, Zeqiu and Wang, Yizhong and Sil, Avirup and Hajishirzi, Hannaneh},
    booktitle = {International Conference on Learning Representations},
    year = {2024},
}

@inproceedings{niu2024ragtruth,
    title = {{RAGT}ruth: A Hallucination Corpus for Developing Trustworthy Retrieval-Augmented Language Models},
    author = {Niu, Cheng and Wu, Yuanhao and Zhu, Juno and Xu, Siliang and Shum, KaShun and Zhong, Randy and Song, Juntong and Zhang, Tong},
    booktitle = {Proceedings of the 62nd Annual Meeting of the Association for Computational Linguistics (Volume 1: Long Papers)},
    pages = {10862--10878},
    year = {2024},
    publisher = {Association for Computational Linguistics},
}

@inproceedings{ayala2024reducing,
    title = {Reducing Hallucination in Structured Outputs via Retrieval-Augmented Generation},
    author = {Ayala, Orlando and Bechard, Patrice},
    booktitle = {Proceedings of the 2024 Conference of the North American Chapter of the Association for Computational Linguistics: Human Language Technologies (Volume 6: Industry Track)},
    pages = {228--238},
    year = {2024},
    publisher = {Association for Computational Linguistics},
}

@inproceedings{thakur2021beir,
    title = {{BEIR}: A Heterogeneous Benchmark for Zero-shot Evaluation of Information Retrieval Models},
    author = {Thakur, Nandan and Reimers, Nils and R{\"u}ckl{\'e}, Andreas and Srivastava, Abhishek and Gurevych, Iryna},
    booktitle = {Proceedings of the Neural Information Processing Systems Track on Datasets and Benchmarks},
    volume = {1},
    year = {2021},
}

@inproceedings{muennighoff2023mteb,
    title = {{MTEB}: Massive Text Embedding Benchmark},
    author = {Muennighoff, Niklas and Tazi, Nouamane and Magne, Loic and Reimers, Nils},
    booktitle = {Proceedings of the 17th Conference of the European Chapter of the Association for Computational Linguistics},
    pages = {2014--2037},
    year = {2023},
    publisher = {Association for Computational Linguistics},
}

@article{zhang2023miracl,
    title = {{MIRACL}: A Multilingual Retrieval Dataset Covering 18 Diverse Languages},
    author = {Zhang, Xinyu and Thakur, Nandan and Ogundepo, Odunayo and Kamalloo, Ehsan and Alfonso-Hermelo, David and Li, Xiaoguang and Liu, Qun and Rezagholizadeh, Mehdi and Lin, Jimmy},
    journal = {Transactions of the Association for Computational Linguistics},
    volume = {11},
    pages = {1114--1131},
    year = {2023},
    publisher = {MIT Press},
}

@inproceedings{zhang2021mrtydi,
    title = {Mr. {TyDi}: A Multi-lingual Benchmark for Dense Retrieval},
    author = {Zhang, Xinyu and Ma, Xueguang and Shi, Peng and Lin, Jimmy},
    booktitle = {Proceedings of the 1st Workshop on Multilingual Representation Learning},
    pages = {127--137},
    year = {2021},
    publisher = {Association for Computational Linguistics},
}

@inproceedings{petroni2021kilt,
    title = {{KILT}: a Benchmark for Knowledge Intensive Language Tasks},
    author = {Petroni, Fabio and Piktus, Aleksandra and Fan, Angela and Lewis, Patrick and Yazdani, Majid and De Cao, Nicola and Thorne, James and Jernite, Yacine and Karpukhin, Vladimir and Maillard, Jean and Plachouras, Vassilis and Rockt{\"a}schel, Tim and Riedel, Sebastian},
    booktitle = {Proceedings of the 2021 Conference of the North American Chapter of the Association for Computational Linguistics: Human Language Technologies},
    pages = {2523--2544},
    year = {2021},
}

@article{daniel2016msmarco,
    title = {{MS} {MARCO}: A Human Generated {MA}chine Reading {CO}mprehension Dataset},
    author = {Daniel Fernando Campos and Tri Nguyen and Mir Rosenberg and Xia Song and Jianfeng Gao and Saurabh Tiwary and Rangan Majumder and Li Deng and Bhaskar Mitra},
    journal = {arXiv preprint arXiv:1611.09268},
    year = {2016},
}

@article{bonifacio2021mmarco,
    title = {{mMARCO}: A Multilingual Version of the {MS} {MARCO} Passage Ranking Dataset},
    author = {Bonifacio, Luiz and Campiotti, Israel and Lotufo, Roberto and Nogueira, Rodrigo},
    journal = {arXiv preprint arXiv:2108.13897},
    year = {2021},
}

@article{bonifacio2022inpars,
    title = {{InPars}: Data Augmentation for Information Retrieval using Large Language Models},
    author = {Bonifacio, Luiz and Jeronymo, Vitor and Abonizio, Hugo and Campiotti, Israel and Lotufo, Roberto and Nogueira, Rodrigo},
    journal = {arXiv preprint arXiv:2202.05144},
    year = {2022},
}

@inproceedings{dai2023promptagator,
    title = {{P}romptagator: Few-shot Dense Retrieval From 8 Examples},
    author = {Dai, Zhuyun and Zhao, Vincent Y. and Ma, Ji and Luan, Yi and Ni, Jianmo and Lu, Jing and Bakalov, Anton and Guu, Kelvin and Hall, Keith B. and Chang, Ming-Wei},
    booktitle = {International Conference on Learning Representations},
    year = {2023},
}

@inproceedings{chaudhary2024relative,
    title = {It's All Relative! -- A Synthetic Query Generation Approach for Improving Zero-Shot Relevance Prediction},
    author = {Chaudhary, Aditi and Raman, Karthik and Bendersky, Michael},
    booktitle = {Findings of the Association for Computational Linguistics: NAACL 2024},
    pages = {1645--1664},
    year = {2024},
    publisher = {Association for Computational Linguistics},
}

@article{zhao2024doclayout,
    title = {{DocLayout-YOLO}: Enhancing Document Layout Analysis through Diverse Synthetic Data and Global-to-Local Adaptive Perception},
    author = {Zhao, Zhiyuan and Kang, Hengrui and Wang, Bin and He, Conghui},
    journal = {arXiv preprint arXiv:2410.12628},
    year = {2024},
}

@inproceedings{chen2024m3embedding,
    title = {{M3}-Embedding: Multi-Linguality, Multi-Functionality, Multi-Granularity Text Embeddings Through Self-Knowledge Distillation},
    author = {Chen, Jianlyu and Xiao, Shitao and Zhang, Peitian and Luo, Kun and Lian, Defu and Liu, Zheng},
    booktitle = {Findings of the Association for Computational Linguistics: ACL 2024},
    pages = {2318--2335},
    year = {2024},
    publisher = {Association for Computational Linguistics},
}

@inproceedings{conneau-etal-2020-unsupervised,
    title = {Unsupervised Cross-lingual Representation Learning at Scale},
    author = {Conneau, Alexis  and
      Khandelwal, Kartikay  and
      Goyal, Naman  and
      Chaudhary, Vishrav  and
      Wenzek, Guillaume  and
      Guzm{\'a}n, Francisco  and
      Grave, Edouard  and
      Ott, Myle  and
      Zettlemoyer, Luke  and
      Stoyanov, Veselin},
    booktitle = {Proceedings of the 58th Annual Meeting of the Association for Computational Linguistics},
    year = {2020},
    publisher = {Association for Computational Linguistics},
    pages = {8440--8451},
}

@article{wang2024multilingual,
    title = {Multilingual {E5} Text Embeddings: A Technical Report},
    author = {Wang, Liang and Yang, Nan and Huang, Xiaolong and Yang, Linjun and Majumder, Rangan and Wei, Furu},
    journal = {arXiv preprint arXiv:2402.05672},
    year = {2024},
}

@article{yu2024arctic,
    title = {{Arctic-Embed} 2.0: Multilingual Retrieval Without Compromise},
    author = {Yu, Puxuan and Merrick, Luke and Nuti, Gaurav and Campos, Daniel},
    journal = {arXiv preprint arXiv:2412.04506},
    year = {2024},
}

@article{babakhin2025llamaembed,
    title = {{Llama-Embed-Nemotron-8B}: A Universal Text Embedding Model for Multilingual and Cross-Lingual Tasks},
    author = {Babakhin, Yauhen and Osmulski, Radek and Ak, Ronay and Moreira, Gabriel and Xu, Mengyao and Schifferer, Benedikt and Liu, Bo and Oldridge, Even},
    journal = {arXiv preprint arXiv:2511.07025},
    year = {2025},
}

@article{zhang2025qwen3embedding,
    title = {{Qwen3} Embedding: Advancing Text Embedding and Reranking Through Foundation Models},
    author = {Zhang, Yanzhao and Li, Mingxin and Long, Dingkun and Zhang, Xin and Lin, Huan and Yang, Baosong and Xie, Pengjun and Yang, An and Liu, Dayiheng and Lin, Junyang and Huang, Fei and Zhou, Jingren},
    journal = {arXiv preprint arXiv:2506.05176},
    year = {2025},
}

@article{vera2025embeddinggemma,
    title = {{EmbeddingGemma}: Powerful and Lightweight Text Representations},
    author = {Vera, Henrique Schechter and Dua, Sahil and Zhang, Biao and Salz, Daniel and Mullins, Ryan and Panyam, Sindhu Raghuram and Smoot, Sara and Naim, Iftekhar and Zou, Joe and Chen, Feiyang and Cer, Daniel and Lisak, Alice and Choi, Min and Gonzalez, Lucas and Sanseviero, Omar and Cameron, Glenn and Ballantyne, Ian and Black, Kat and Chen, Kaifeng and Wang, Weiyi and Li, Zhe and Martins, Gus and Lee, Jinhyuk and Sherwood, Mark and Ji, Juyeong and Wu, Renjie and Zheng, Jingxiao and Singh, Jyotinder and Sharma, Abheesht and Sreepathihalli, Divyashree and Jain, Aashi and Elarabawy, Adham and Co, AJ and Doumanoglou, Andreas and Samari, Babak and Hora, Ben and Potetz, Brian and Kim, Dahun and Alfonseca, Enrique and Moiseev, Fedor and Han, Feng and Gomez, Frank Palma and Ábrego, Gustavo Hernández and Zhang, Hesen and Hui, Hui and Han, Jay and Gill, Karan and Chen, Ke and Chen, Koert and Shanbhogue, Madhuri and Boratko, Michael and Suganthan, Paul and Duddu, Sai Meher Karthik and Mariserla, Sandeep and Ariafar, Setareh and Zhang, Shanfeng and Zhang, Shijie and Baumgartner, Simon and Goenka, Sonam and Qiu, Steve and Dabral, Tanmaya and Walker, Trevor and Rao, Vikram and Khawaja, Waleed and Zhou, Wenlei and Ren, Xiaoqi and Xia, Ye and Chen, Yichang and Chen, Yi-Ting and Dong, Zhe and Ding, Zhongli and Visin, Francesco and Liu, Gaël and Zhang, Jiageng and Kenealy, Kathleen and Casbon, Michelle and Kumar, Ravin and Mesnard, Thomas and Gleicher, Zach and Brick, Cormac and Lacombe, Olivier and Roberts, Adam and Yin, Qin and Sung, Yunhsuan and Hoffmann, Raphael and Warkentin, Tris and Joulin, Armand and Duerig, Tom and Seyedhosseini, Mojtaba},
    journal = {arXiv preprint arXiv:2509.20354},
    year = {2025},
}

@inproceedings{zhang2024mgte,
    title = {{mGTE}: Generalized Long-Context Text Representation and Reranking Models for Multilingual Text Retrieval},
    author = {Zhang, Xin and Zhang, Yanzhao and Long, Dingkun and Xie, Wen and Dai, Ziqi and Tang, Jialong and Lin, Huan and Yang, Baosong and Xie, Pengjun and Huang, Fei and Zhang, Meishan and Li, Wenjie and Zhang, Min},
    booktitle = {Proceedings of the 2024 Conference on Empirical Methods in Natural Language Processing: Industry Track},
    pages = {1393--1412},
    year = {2024},
    publisher = {Association for Computational Linguistics},
}

@misc{meta2025llama4,
    title = {Introducing {LLaMA} 4: Advancing Multimodal Intelligence},
    author = {{Meta AI}},
    year = {2025},
    howpublished = {\url{https://ai.meta.com/blog/llama-4-multimodal-intelligence/}},
    note = {Accessed: 2025-11-28}
}

@article{yang2025qwen3,
    title = {{Qwen3} Technical Report},
    author = {Yang, An and Li, Anfeng and Yang, Baosong and Zhang, Beichen and Hui, Binyuan and Zheng, Bo and Yu, Bowen and Gao, Chang and Huang, Chengen and Lv, Chenxu and Zheng, Chujie and Liu, Dayiheng and Zhou, Fan and Huang, Fei and Hu, Feng and Ge, Hao and Wei, Haoran and Lin, Huan and Tang, Jialong and Yang, Jian and Tu, Jianhong and Zhang, Jianwei and Yang, Jianxin and Yang, Jiaxi and Zhou, Jing and Zhou, Jingren and Lin, Junyang and Dang, Kai and Bao, Keqin and Yang, Kexin and Yu, Le and Deng, Lianghao and Li, Mei and Xue, Mingfeng and Li, Mingze and Zhang, Pei and Wang, Peng and Zhu, Qin and Men, Rui and Gao, Ruize and Liu, Shixuan and Luo, Shuang and Li, Tianhao and Tang, Tianyi and Yin, Wenbiao and Ren, Xingzhang and Wang, Xinyu and Zhang, Xinyu and Ren, Xuancheng and Fan, Yang and Su, Yang and Zhang, Yichang and Zhang, Yinger and Wan, Yu and Liu, Yuqiong and Wang, Zekun and Cui, Zeyu and Zhang, Zhenru and Zhou, Zhipeng and Qiu, Zihan},
    journal = {arXiv preprint arXiv:2505.09388},
    year = {2025},
}

@inproceedings{gao2024enhancing,
    title = {Enhancing Legal Case Retrieval via Scaling High-quality Synthetic Query-Candidate Pairs},
    author = {Gao, Cheng and Xiao, Chaojun and Liu, Zhenghao and Chen, Huimin and Liu, Zhiyuan and Sun, Maosong},
    booktitle = {Proceedings of the 2024 Conference on Empirical Methods in Natural Language Processing},
    pages = {7086--7100},
    year = {2024},
    publisher = {Association for Computational Linguistics},
}

@inproceedings{oderinde2025aviation,
    title = {Aviation Safety {QA} Dataset for Extracting Knowledge From Incident Reports},
    author = {Oderinde, Timilehin P. and Chandra, Chetan and Albertoli, Leslie and Bhanpato, Jirat and Bendarkar, Mayank V. and Mavris, Dimitri},
    booktitle = {AIAA AVIATION FORUM AND ASCEND 2025},
    year = {2025},
    doi = {10.2514/6.2025-3248},
}

@article{amiri2025chunktwice,
    title = {Chunk Twice, Embed Once: A Systematic Study of Segmentation and Representation Trade-offs in Chemistry-Aware Retrieval-Augmented Generation},
    author = {Amiri, Mahmoud and Bocklitz, Thomas},
    journal = {arXiv preprint arXiv:2506.17277},
    year = {2025},
}

@inproceedings{rekabsaz2021tripclick,
    title = {{TripClick}: The Log Files of a Large Health Web Search Engine},
    author = {Rekabsaz, Navid and Lesota, Oleg and Schedl, Markus and Brassey, Jon and Eickhoff, Carsten},
    booktitle = {Proceedings of the 44th International ACM SIGIR Conference on Research and Development in Information Retrieval},
    pages = {2507--2513},
    year = {2021},
    doi = {10.1145/3404835.3463242}
}

@inproceedings{robertson1995okapi,
    title = {Okapi at {TREC}-3},
    author = {Robertson, Stephen E. and Walker, Steve and Jones, Susan and Hancock-Beaulieu, Micheline and Gatford, Mike},
    booktitle = {Overview of the Third Text REtrieval Conference (TREC-3)},
    pages = {109--126},
    year = {1995},
    publisher = {NIST Special Publication 500-225}
}

@misc{langchain2022,
    title = {LangChain: Building applications with {LLMs} through composability},
    author = {Chase, Harrison},
    year = {2022},
    url = {https://github.com/langchain-ai/langchain},
    note = {Open source software}
}

@article{honnibal2017spacy,
    title = {{spaCy 2}: Natural language understanding with {B}loom embeddings, convolutional neural networks and incremental parsing},
    author = {Honnibal, Matthew and Montani, Ines},
    year = {2017},
    journal = {To appear},
    url = {https://github.com/explosion/spaCy}
}

@article{nelson1995nasa,
    title = {The {NASA} technical report server},
    author = {Nelson, Michael L. and Gottlich, Gretchen L. and Bianco, David J. and Paulson, Sharon S. and Binkley, Robert L. and Kellogg, Yvonne D. and Beaumont, Chris J. and Schmunk, Robert B. and Kurtz, Michael J. and Accomazzi, Alberto and Syed, Omar},
    journal = {Internet Research},
    volume = {5},
    number = {2},
    pages = {25--36},
    year = {1995},
    doi = {10.1108/10662249510094768}
}

@inproceedings{emmons2024text,
    title = {Text Summarization in Aviation Safety: A Comparative Study of Large Language Models},
    author = {Emmons, Jonathan and Sharma, Taneesha and Matthews, Bryan and Salloum, Mariam},
    booktitle = {AIAA Aviation Forum 2024},
    year = {2024},
    doi = {10.2514/6.2024-4569}
}

@inproceedings{liu2023geval,
    title = {{G}-Eval: {NLG} Evaluation using Gpt-4 with Better Human Alignment},
    author = {Liu, Yang  and
      Iter, Dan  and
      Xu, Yichong  and
      Wang, Shuohang  and
      Xu, Ruochen  and
      Zhu, Chenguang},
    booktitle = {Proceedings of the 2023 Conference on Empirical Methods in Natural Language Processing},
    year = {2023},
}

@inproceedings{liu2023robustness,
    title = {On the Robustness of Generative Retrieval Models: An Out-of-Distribution Perspective},
    author = {Liu, Yu-An and Zhang, Ruqing and Guo, Jiafeng and Chen, Wei and Cheng, Xueqi},
    booktitle = {Proceedings of the Gen-IR Workshop at SIGIR 2023},
    year = {2023},
}

@article{voorhees2021treccovid,
    title = {{TREC-COVID}: Constructing a Pandemic Information Retrieval Test Collection},
    author = {Voorhees, Ellen and Alam, Tasmeer and Bedrick, Steven and Demner-Fushman, Dina and Hersh, William R. and Lo, Kyle and Roberts, Kirk and Soboroff, Ian and Wang, Lucy Lu},
    journal = {ACM SIGIR Forum},
    volume = {54},
    number = {1},
    pages = {1--12},
    year = {2021},
}

@inproceedings{adams2023sparse,
    title = {From Sparse to Dense: {GPT}-4 Summarization with Chain of Density Prompting},
    author = {Adams, Griffin and Fabbri, Alexander R. and Ladhak, Faisal and Lehman, Eric and Elhadad, No{\'e}mie},
    booktitle = {Proceedings of the 4th New Frontiers in Summarization Workshop},
    pages = {68--74},
    year = {2023},
    publisher = {Association for Computational Linguistics},
}

@misc{meng2024sfr,
    title = {{SFR}-Embedding-Mistral: Enhance Text Retrieval with Transfer Learning},
    author = {Meng, Rui and Liu, Ye and Joty, Shafiq Rayhan and Xiong, Caiming and Zhou, Yingbo and Yavuz, Semih},
    howpublished = {Salesforce AI Research Blog},
    year = {2024},
    url = {https://www.salesforce.com/blog/sfr-embedding/}
}

@misc{deepseekai2025v32exp,
    title = {{DeepSeek-V3.2-Exp}: Boosting Long-Context Efficiency with {DeepSeek} Sparse Attention},
    author = {DeepSeek-AI},
    year = {2025},
    howpublished = {\url{https://github.com/deepseek-ai/DeepSeek-V3.2-Exp}},
    note = {Accessed: 2025-10-15}
}

@misc{robyn_speer_2022_7199437,
  author       = {Speer, Robyn},
  title        = {rspeer/wordfreq: v3.0},
  month        = sep,
  year         = 2022,
  publisher    = {Zenodo},
  version      = {v3.0.2},
  doi          = {10.5281/zenodo.7199437},
}

@article{meta2024llama3herd,
  title={The Llama 3 Herd of Models},
  author={Meta AI},
  journal={arXiv preprint arXiv:2407.21783},
  year={2024},
  note={Definitive technical report for the Llama 3.1 family, including the 8B Instruct model.}
}

@article{jiang2023mistral7b,
  title={Mistral 7B},
  author = {Jiang, Albert Q. and Sablayrolles, Alexandre and Mensch, Arthur and Bamford, Chris and Chaplot, Devendra Singh and de las Casas, Diego and Bressand, Florian and Lengyel, Gianna and Lample, Guillaume and Saulnier, Lucile and Renard Lavaud, Lélio and Lachaux, Marie-Anne and Stock, Pierre and Le Scao, Teven and Lavril, Thibaut and Wang, Thomas and Lacroix, Timothée and El Sayed, William},
  journal={arXiv preprint arXiv:2310.06825},
  year={2023},
}

@misc{deepseekai2025deepseekr1,
  title={DeepSeek-R1: Incentivizing Reasoning Capability in LLMs via Reinforcement Learning},
  author={DeepSeek-AI},
  year={2025},
  journal={arXiv preprint arXiv:2501.12948},
}

@inproceedings{madaan2023selfrefine,
    title = {SELF-REFINE: iterative refinement with self-feedback},
    author = {Madaan, Aman and Tandon, Niket and Gupta, Prakhar and Hallinan, Skyler and Gao, Luyu and Wiegreffe, Sarah and Alon, Uri and Dziri, Nouha and Prabhumoye, Shrimai and Yang, Yiming and Gupta, Shashank and Majumder, Bodhisattwa Prasad and Hermann, Katherine and Welleck, Sean and Yazdanbakhsh, Amir and Clark, Peter},
    booktitle = {Proceedings of the 37th International Conference on Neural Information Processing Systems},
    pages = {61},
    year = {2023},
}

@inproceedings{shinn2023reflexion,
    title = {Reflexion: language agents with verbal reinforcement learning},
    author = {Shinn, Noah and Cassano, Federico and Gopinath, Ashwin and Narasimhan, Karthik and Yao, Shunyu},
    booktitle = {Proceedings of the 37th International Conference on Neural Information Processing Systems},
    pages = {19},
    year = {2023},
}

@inproceedings{wang2025critique,
    title = {Self-Critique and Refinement for Faithful Natural Language Explanations},
    author = {Wang, Yingming  and
      Atanasova, Pepa},
    booktitle = {Proceedings of the 2025 Conference on Empirical Methods in Natural Language Processing},
    pages = {8492--8518},
    year = {2025},
}






\appendix

\section{Prompt Engineering Details}
\label{sec:appendix_prompts}

This section provides the detailed prompt templates used in the construction of the STELLA benchmark. All prompts were designed for and executed on GPT-5. In the templates below, placeholders enclosed in angle brackets (e.g., \texttt{<variable>}) indicate where dynamic data is inserted at runtime.

\subsection{Query Intent Classification Prompts}
\label{sec:appendix_intent_prompts}
As described in Section~\ref{sec:query_intent_classification}, we utilized GPT-5 to classify candidate passages into five distinct query intents. Table~\ref{tab:prompt_intent_classification} presents the exact prompt template derived from our pipeline.

\begin{table*}[h]
    \centering
    \small
    \begin{tabular}{p{0.95\textwidth}}
        \toprule
        \textbf{Query Intent Classification Prompt Template} \\
        \midrule
        \textbf{\# Role}\\ You are an aerospace domain expert. Classify the passage into the single best query intent for retrieval. \\
        \vspace{0.1em}
        Below are 5 query intents for the aerospace domain: \\
        - Definition / Principle: distinguish concepts, explain mechanisms of operation, approximations, or variable dependencies. \\
        - Numerical / Specification: retrieve numeric/spec details: values, ranges, assumptions, units, uncertainties (avoid single-number lookup). \\
        - Procedure / Operation: steps, initialization/calibration, scheduling, operational rules. \\
        - Comparison / Trade-off: quantitative comparison of performance, mass, power, margins across options/configurations. \\
        - Anomaly / Risk: causes of anomalies/failures, reproduction conditions, mitigations/recurrence prevention. \\
        \vspace{1em}
        Which of the 5 intents is the following passage best suited to generate a query for? \\
        You must answer only with one of the 5 intent names. (e.g., Definition / Principle) \\
        \vspace{1em}
        Passage: \\
        \texttt{<passage\_text>} \\
        \vspace{1em}
        Most Suitable Intent: \\
        \bottomrule
    \end{tabular}
    \caption{Prompt template used for Query Intent Classification. The \texttt{<passage\_text>} placeholder is replaced with the actual text of the candidate passage at runtime.}
    \label{tab:prompt_intent_classification}
\end{table*}

\subsection{Synthetic Query Generation Prompts}
\label{sec:appendix_generation_prompts}
This section details the prompt templates used for the three stages of synthetic query generation: (1) terminology description generation, (2) Terminology Concordant Query  Generation (TCQG), and (3) Terminology Agnostic Query Generation (TAQG).

\subsubsection{Terminology Description Generation}
\label{sec:appendix_tdg}
Table~\ref{tab:prompt_term_desc} shows the prompt used to generate context-aware descriptions for technical terms. These descriptions are crucial for the TAQG process, where they serve as semantic substitutes for the actual terms.

\begin{table*}[h]
    \centering
    \small
    \begin{tabular}{p{0.95\textwidth}}
        \toprule
        \textbf{Terminology Description Generation Prompt Template} \\
        \midrule
        \textbf{\# Role}\\
        You are an aerospace domain expert. The following is a context excerpted from a specific document. Based on the context, generate a short and clear description for the technical term "\texttt{<term>}". If the meaning cannot be determined from the context, answer "Difficult to define within context".\\
        \vspace{0.5em}
        Context: \\
        \texttt{<context\_text>} \\
        \vspace{0.5em}
        Term:\\ \texttt{<term>} \\
        \vspace{0.5em}
        Short Description: \\
        \bottomrule
    \end{tabular}
    \caption{Prompt template for generating context-aware terminology descriptions.}
    \label{tab:prompt_term_desc}
\end{table*}

\clearpage

\subsubsection{Terminology Concordant Query Generation (TCQG)}
Table~\ref{tab:prompt_tcq} presents the prompt used for generating TCQs. This prompt implements the Chain-of-Density (CoD) process and \textsc{Self-Reflection} while explicitly enforcing the inclusion of terminology found in the passage.

\begin{table*}[h]
    \centering
    \scriptsize 
    \begin{tabular}{p{0.95\textwidth}}
        \toprule
        \textbf{TCQG Prompt Template} \\
        \midrule
        \vspace{0.1em}
        \textbf{\# Role} \\
        You generate synthetic queries for an aerospace IR benchmark. Apply a three-step Chain-of-Density (CoD) process to progressively densify entities from the Passage. At each step, list which entities you recognized/added and provide one line of English self-feedback that improves the next step. \\
        \vspace{0.1em}
        \textbf{\# Entity Definition} \\
        An "entity" must be a short, specific technical term or noun phrase (1-3 words). Avoid: Do not use long, descriptive phrases or full clauses. \\
        \vspace{0.1em}
        \textbf{\# Inputs (you will receive exactly one JSON object)} \\
        \texttt{<input\_json>} \\
        \# where: \\
        \# - passage\_text: raw text (single passage) \\
        \# - identified\_terms: [\{ "term": "<tech term (EN)>", "description": "<short note>" \}...] \\
        \# - sampled\_intention: exactly one of: \\\#\hspace{1em} - Definition / Principle — distinguish concepts, explain mechanisms/approximations/variable dependencies\\\#\hspace{1em} - Numerical / Specification — values, ranges, assumptions, units, uncertainties (avoid single-number lookup)\\\#\hspace{1em} - Procedure / Operation — steps, initialization/calibration, schedules, operational rules\\\#\hspace{1em} - Comparison / Trade-off — quantitative comparisons of performance/mass/power/margins across options/configs\\\#\hspace{1em} - Anomaly / Risk — causes, reproduction conditions, mitigations/recurrence prevention \\
        \vspace{0.1em}
        \textbf{\# CoD Steps} \\
        - Step 1 (Seed): Intentionally omit some core entities and write one-sentence query answerable from the Passage alone, while STRICTLY avoiding all terms in \texttt{identified\_terms.term}. \\
        - Step 2 (Densify-1): Expand Step 1 by adding one term from \texttt{identified\_terms}. term verbatim (keep English spelling). The "description" field is only for your understanding—do not copy it. \\
        - Step 3 (Densify-2): Refine Step 2 by adding one term from \texttt{identified\_terms}. term verbatim (keep English spelling). The "description" field is only for your understanding—do not copy it. \\
        \vspace{0.1em}
        \textbf{\# Each step must include} \\
        - query: one English sentence (15–25 tokens), neutral/technical tone \\
        - recognized\_entities: array of short Passage entities (terms/noun phrases) reflected in this step’s query (MAXIMUM 2 entities) \\
        - entities\_added: array of new entities added vs. the previous step (Step 1 may be empty) \\
        - self\_feedback: one concise English line describing actionable improvements for the next step\\
        \vspace{0.1em}
        \textbf{\# Self-Feedback Guidance (every step)} \\
        Write one compact, imperative line (semicolons to chain 2–3 items). Always include: \\
        1) Intention adherence to "sampled\_intention"; \\
        2) Constraint checks: passage-only, avoid forbidden forms, 15–25 tokens, no outside knowledge; \\
        3) Next actions: name specific entities/conditions to add/refine (e.g., mechanism factor, operating condition, range/assumption/unit, comparison metric).\\
        \vspace{0.1em}
        \textbf{\# Hard Constraints (all steps)} \\
        1) Passage-answerable only (safe inference). \\
        2) Forbidden forms: no list/quote requests; no single-number lookup; no yes/no prompts; no bare deictics (``this/that/these/those''). \\
        3) Style/Length: exactly one sentence in English; neutral, technical; 15–25 tokens; use Passage terms or safe synonyms only. \\
        4) Intention preservation: never change "sampled\_intention". \\
        5) No external knowledge: no outside facts/sources/tool names; no invented symbols/variables. \\
        6) Tech term use: Step 2 and Step 3 must include one \texttt{identified\_terms.term} verbatim (EN). Do not copy any description. \\
        7) Entity Granularity: All items in "recognized\_entities" and "entities\_added" MUST adhere to the \# Entity Definition (i.e., short, specific terms or noun phrases, 1-3 words). Never list long clauses or full sentences as entities. \\
        8) "identified\_terms" Reservation: Do NOT use any term from the "identified\_terms" list in the "query", "recognized\_entities", or "entities\_added" fields for Step 1. These terms are reserved exclusively for introduction in Step 2 and Step 3. \\
        \vspace{0.1em}
        \textbf{\# Output Format (return one JSON object only; no extra text/comments/markdown)} \\
        \{ \\
        \hspace{1em} "intention": "<copy sampled\_intention>", \\
        \hspace{1em} "step\_1": \{ \\
        \hspace{1em} "query": "<one English sentence>", \\
        \hspace{1em} "recognized\_entities": ["<entity>", "..."], \\
        \hspace{1em} "entities\_added": [], \\
        \hspace{1em} "self\_feedback": "<one English line>" \\
        \hspace{1em} \}, \\
        \hspace{1em} "step\_2": \{ \\
        \hspace{1em} "query": "<one English sentence including one identified term verbatim>", \\
        \hspace{1em} "recognized\_entities": ["<entity>", "..."], \\
        \hspace{1em} "entities\_added": ["<one entity from identified\_terms.term>"], \\
        \hspace{1em} "self\_feedback": "<one English line>" \\
        \hspace{1em} \}, \\
        \hspace{1em} "step\_3": \{ \\
        \hspace{1em} "query": "<one English sentence including one identified term verbatim>", \\
        \hspace{1em} "recognized\_entities": ["<entity>", "..."], \\
        \hspace{1em} "entities\_added": ["<one entity from identified\_terms.term>"], \\
        \hspace{1em} "self\_feedback": "<one English line>" \\
        \hspace{1em} \} \\
        \} \\
        \bottomrule
    \end{tabular}
    \caption{Instruction prompt for TCQG using CoD and \textsc{Self-Reflection}.}
    \label{tab:prompt_tcq}
\end{table*}

\clearpage

\subsubsection{Terminology Agnostic Query Generation (TAQG)}
Table~\ref{tab:prompt_taq} shows the prompt for TAQG. This prompt is characterized by an \textbf{Absolute Term-Ban Policy} that strictly prohibits the usage of specific terms, requiring the model to rely on indirect descriptions instead.

\begin{table*}[h]
    \centering
    \scriptsize 
    \begin{tabular}{p{0.95\textwidth}}
        \toprule
        \textbf{TAQG Prompt Template} \\
        \midrule
        \vspace{0.1em}
        \textbf{\# Role} \\
        \textit{(Note: identical to the TCQG prompt shown in Table~\ref{tab:prompt_tcq}.)} \\
        \vspace{0.1em}
        \textbf{\# Entity Definition} \\
        An "entity" must be a short, specific technical term or noun phrase (1-3 words). Avoid: Do not use long, descriptive phrases or full clauses. \\
        \vspace{0.1em}
        \textbf{\# ABSOLUTE TERM-BAN POLICY} \\
        You MUST NOT include ANY token that matches any \texttt{identified\_terms.term} in any step's query — regardless of case, hyphenation, pluralization, or spelling variants. Instead, convey the concept indirectly using concise paraphrases derived from the corresponding "description". Do NOT copy the "description" verbatim; paraphrase it and avoid quotation marks. \\
        \vspace{0.1em}
        \textbf{\# Inputs (you will receive exactly one JSON object)} \\
        \textit{(Note: identical to the TCQG prompt shown in Table~\ref{tab:prompt_tcq}.)} \\
        \vspace{0.1em}
        \textbf{\# CoD Steps} \\
        - Step 1 (Seed): Intentionally omit some core entities and write one-sentence query answerable from the Passage alone, while STRICTLY avoiding all terms in \texttt{identified\_terms.term}. \\
        - Step 2 (Densify-1): Expand Step 1 by INDIRECTLY encode one concept from "identified\_terms" by paraphrasing its "description" (no verbatim copy, no quotes). Continue to avoid ALL terms in \texttt{identified\_terms.term}. \\
        - Step 3 (Densify-2): Refine Step 2 by INDIRECTLY encode one concept from "identified\_terms" by paraphrasing its "description" (no verbatim copy, no quotes). Continue to avoid ALL terms in \texttt{identified\_terms.term}. \\
        \vspace{0.1em}
        \textbf{\# Each step must include} \\
        - query: one English sentence (15–25 tokens), neutral/technical tone, containing zero tokens equal to any "identified\_term" \\
        - recognized\_entities: array of short Passage entities (terms/noun phrases) reflected in this step’s query (MAXIMUM 2 entities) \\
        - entities\_added: array of new entities added vs. the previous step (Step 1 may be empty) \\
        - self\_feedback: one concise English line describing actionable improvements for the next step\\
        - (Step 2, Step 3) "descriptions\_referenced": array listing which descriptions (short paraphrase tags or brief identifiers) you leveraged to indirectly encode the concept (no verbatim copying) \\
        \vspace{0.1em}
        \textbf{\# Self-Feedback Guidance (every step)} \\
        Write one compact, imperative line (semicolons to chain 2–3 items). Always include: \\
        1) Intention adherence to "sampled\_intention"; \\
        2) Constraint checks: passage-only, avoid forbidden forms, 15–25 tokens, zero forbidden-term overlap, no outside knowledge; \\
        3) Next actions: name specific entities/conditions to add/refine using indirect phrasing from descriptions (e.g., mechanism factor, operating condition, range/assumption/unit, comparison metric).\\
        \vspace{0.1em}
        \textbf{\# Hard Constraints (all steps)} \\
        1) Passage-answerable only (safe inference). \\
        2) Forbidden forms: no list/quote requests; no single-number lookup; no yes/no prompts; no bare deictics (``this/that/these/those''). \\
        3) Style/Length: exactly one sentence in English; neutral, technical; 15–25 tokens; use Passage terms or safe synonyms only; never use any identified term. \\
        4) Intention preservation: never change "sampled\_intention". \\
        5) No external knowledge: no outside facts/sources/tool names; no invented symbols/variables. \\
        6) Indirect encoding requirement: In Step 2 and Step 3, paraphrase one description to convey the concept; record it in "descriptions\_referenced" (do not quote or copy). \\
        7) Entity Granularity: All items in "recognized\_entities" and "entities\_added" MUST adhere to the \# Entity Definition (i.e., short, specific terms or noun phrases, 1-3 words). Never list long clauses or full sentences as entities. \\
        8) "identified\_terms" Reservation: Do NOT use any term from the "identified\_terms" list in the "query", "recognized\_entities", or "entities\_added" fields for Step 1. These terms are reserved exclusively for introduction in Step 2 and Step 3. \\
        \vspace{0.1em}
        \textbf{\# Output Format (return one JSON object only; no extra text/comments/markdown)} \\
        \{ \\
        \hspace{1em} "intention": "<copy sampled\_intention>", \\
        \hspace{1em} "step\_1": \{ \\
        \hspace{1em} "query": "<one English sentence (no identified\_terms)>", \\
        \hspace{1em} "recognized\_entities": ["<entity>", "..."], \\
        \hspace{1em} "entities\_added": [], \\
        \hspace{1em} "self\_feedback": "<one English line>" \\
        \hspace{1em} \}, \\
        \hspace{1em} "step\_2": \{ \\
        \hspace{1em} "query": "<one English sentence (no identified\_terms) with at least one concept encoded via paraphrased description>", \\
        \hspace{1em} "recognized\_entities": ["<entity>", "..."], \\
        \hspace{1em} "entities\_added": ["<one entity from identified\_terms.term>"], \\
        \hspace{1em} "descriptions\_referenced": ["<"term" the exact identified\_terms.term whose concept was indirectly encoded>", "<brief paraphrase tag(s) for which description(s) were leveraged>"], \\
        \hspace{1em} "self\_feedback": "<one English line>" \\
        \hspace{1em} \}, \\
        \hspace{1em} "step\_3": \{ \\
        \hspace{1em} "query": "<one English sentence (no identified\_terms) with at least one concept encoded via paraphrased description>", \\
        \hspace{1em} "recognized\_entities": ["<entity>", "..."], \\
        \hspace{1em} "entities\_added": ["<one entity from identified\_terms.term>"], \\
        \hspace{1em} "descriptions\_referenced": ["<"term" the exact identified\_terms.term whose concept was indirectly encoded>", "<brief paraphrase tag(s) for which description(s) were leveraged>"], \\
        \hspace{1em} "self\_feedback": "<one English line>" \\
        \hspace{1em} \} \\
        \} \\
        \bottomrule
    \end{tabular}
    \caption{Instruction prompt for TAQG. Note the addition of the term-ban policy and description referencing fields.}
    \label{tab:prompt_taq}
\end{table*}

\subsection{Cross-lingual Translation Prompts}
\label{sec:appendix_translation_prompts}
To construct the cross-lingual evaluation sets described in Section~\ref{sec:cross_lingual_extension}, we utilized a dynamic prompting strategy that adapts to the query type. The translation prompt is assembled at runtime depending on whether the target query is a TCQ or a TAQ.

For TCQ translation, the prompt injects a \textbf{Critical Rule} block that explicitly lists the technical terms extracted from the input query (using the dictionary from Section~\ref{sec:terminology_extraction}) and instructs the model to preserve them in English. For TAQ translation, this rule is omitted, allowing for full natural translation. Additionally, we provided language-specific few-shot examples to ensure the model adheres to the desired output format and style. Table~\ref{tab:prompt_translation} details the prompt template and the conditional logic used.

\begin{table*}[h]
    \centering
    \small
    \begin{tabular}{p{0.95\textwidth}}
        \toprule
        \textbf{Translation Prompt Template} \\
        \midrule
        \textbf{\# Role} \\
        You are an expert translator specializing in technical aerospace queries. \\
        Your task is to translate the given English text into \texttt{<target\_language\_name>}. \\
        Follow the rules and examples below precisely. \\
        \vspace{0.5em}
        \textbf{\# Examples} \\
        \texttt{<few\_shot\_examples>} \\
        \textit{(Note: Language-specific examples demonstrating input/output pairs are inserted here. e.g., for Korean: Input: "Define how CFD...", Output: "CFD가...")} \\
        \vspace{0.5em}
        \textbf{\# Rule} \\
        1. Translate the text naturally into \texttt{<target\_language\_name>}. \\
        2. \texttt{<keep\_terms\_instruction>} \\
        3. Provide ONLY the final translated text, with no explanations or conversational text. \\
        \vspace{1em}
        English Text: \\
        \texttt{<input\_query>}\\
        \vspace{1em}
        \textbf{--- Dynamic Instruction Logic (\texttt{<keep\_terms\_instruction>}) ---} \\
        \vspace{0.25em}
        \textit{Case A: TCQ (Hybrid Translation - Terminology Preservation)} \\
        If specific technical terms are detected in the input text: \\
        
        \vspace{0.3em}
        \hspace{1.5em}\parbox{0.9\textwidth}{
            \textbf{CRITICAL RULE FOR THIS REQUEST:} \\
            You MUST NOT translate the following specific technical terms found in this input. \\
            Keep them in their original English form: [\texttt{"term1"}, \texttt{"term2"}, ...].
        }\\
        \vspace{0.3em}\\
        
        \textit{Case B: TAQ (Full Translation)} \\
        If no specific terminology constraints apply: \\
        
        \vspace{0.3em}
        \hspace{1.5em}\parbox{0.9\textwidth}{
            No specific terms to keep for this request.
        }
        \vspace{0.1em}

    \end{tabular}
    \caption{Prompt template for Cross-lingual Extension. The \texttt{<keep\_terms\_instruction>} is dynamically populated based on whether the query is a TCQ or a TAQ, implementing the hybrid translation strategy.}
    \label{tab:prompt_translation}
\end{table*}

\end{document}